\documentclass[12pt]{article}
 
\usepackage{axodraw}

\def\Fint{\rlap{$\Biggl\rfloor$}\Biggl\lceil}

\begin{document}

\begin{titlepage}
 
\begin{flushright}
gr-qc/yymmnnn \\ UFIFT-QG-06-03
\end{flushright}

\vspace{2cm}

\begin{center}
{\bf Leading Log Solution for Inflationary Yukawa}
\end{center}

\vspace{.5cm}

\begin{center}
S. P. Miao$^{\dagger}$ and R. P. Woodard$^{\ddagger}$
\end{center}

\vspace{.5cm}

\begin{center}
\it{Department of Physics \\ 
University of Florida \\
Gainesville, FL 32611}
\end{center}

\vspace{1cm}

\begin{center}
ABSTRACT
\end{center}
We generalize Starobinski\u{\i}'s stochastic technique to the
theory of a massless, minimally coupled scalar interacting with
a massless fermion in a locally de Sitter geometry. The scalar is 
an ``active'' field that can engender infrared logarithms. 
The fermion is a ``passive'' field that cannot cause infrared 
logarithms but which can carry them, and which can also induce 
new interactions between the active fields. The procedure for 
dealing with passive fields is to integrate them out, then 
stochastically simplify the resulting effective action following 
Starobinski\u{\i}. Because Yukawa theory is quadratic in the fermion 
this can be done explicitly using the classic solution of Candelas
and Raine. We check the resulting stochastic formulation against an
explicit two loop computation. We also derive a nonperturbative, 
leading log result for the stress tensor. Because the scalar 
effective potential induced by fermions is unbounded below,
back-reaction from this model might dynamically cancel an arbitrarily
large cosmological constant.

\begin{flushleft}
\end{flushleft}

\begin{flushleft}
$^{\dagger}$ e-mail: miao@phys.ufl.edu \\
$^{\ddagger}$ e-mail: woodard@phys.ufl.edu
\end{flushleft}

\end{titlepage}

\section{Introduction}

Massless, minimally coupled scalars and gravitons are unique in achieving
zero mass without classical conformal invariance. This means that inflation
rips their virtual quanta out of the vacuum, which greatly strengthens the
quantum loop effects they mediate \cite{RPW1}. In the expectation values of 
familiar operators these enhanced quantum effects typically manifest as 
{\it infrared logarithms}. A simple example is provided by the stress tensor 
of a massless, minimally coupled scalar with a quartic self-interaction,
\begin{equation}
\mathcal{L} = -\frac12 \partial_{\mu} \varphi \partial_{\nu} \varphi g^{\mu\nu}
\sqrt{-g} - \frac{\lambda}{4!} \varphi^4 \sqrt{-g} + {\rm counterterms} \; .
\end{equation}
When the expectation value of the stress tensor of this theory is computed
in de Sitter background,
\begin{equation}
ds^2 = -dt^2 + a^2(t) d\vec{x} \cdot d\vec{x} \qquad {\rm with} \qquad
a(t) = e^{H t} \; ,
\end{equation}
and renormalized so as to make quantum effects vanish at $t=0$, the results 
for the quantum-induced energy density and pressure are \cite{OW1,OW2},
\begin{eqnarray}
\rho(t) & = & \frac{\lambda H^4}{(2 \pi)^4} \Biggl\{ \frac18 \ln^2(a) \Biggr\} 
+ O(\lambda^2) \; , \label{rho} \\
p(t) & = & \frac{\lambda H^4}{(2 \pi)^4} \Biggl\{-\frac18 \ln^2(a) - \frac1{12}
\ln(a) \Biggr\} + O(\lambda^2) \; . \label{pres}
\end{eqnarray}
Infrared logarithms are the factors of $\ln(a) = H t$. They arise from the
fact that inflationary particle production drives the free scalar field 
strength away from zero \cite{VF,L,S}
\begin{equation}
\Bigl\langle \Omega \Bigl\vert \varphi^2(x) \Bigr\vert \Omega \Bigr\rangle_0 = 
\frac{H^2}{4 \pi^2} \ln(a) + {\rm Divergent\ Constant} \; . \label{lowest}
\end{equation}
This increases the vacuum energy contributed by the quartic potential and 
the result is evident in (\ref{rho}-\ref{pres}).

Infrared logarithms arise in the one particle irreducible (1PI) functions of
this theory \cite{BOW}. They occur as well in massless, minimally coupled 
scalar quantum electrodynamics (SQED) \cite{PTW1,PTW2,PW1,PW2} and in
massless Yukawa theory \cite{PW3,GP}. The 1PI functions of pure gravity on
de Sitter background show infrared logarithms \cite{TW1,TW2,TW3}. A recent 
all orders analysis of scalar-driven inflation was unable to exclude the 
possibility that they might even contaminate loop corrections to the power 
spectra of cosmological perturbations \cite{SW}! And infrared logarithms
have been discovered in the 1PI functions of Gravity + Dirac \cite{MW1,MW2}.

Infrared logarithms are fascinating because they introduce a secular element
into the usual, static expansion in the loop counting parameter. No matter
how small the coupling constant $\lambda$ is in (\ref{rho}-\ref{pres}), the
continued growth of the inflationary scale factor must eventually overwhelm 
it. When this happens perturbation theory typically breaks down. For example,
the general form of the induced energy density (\ref{rho}) is,
\begin{equation}
\rho(t) = H^4 \sum_{\ell = 2}^{\infty} \lambda^{\ell-1} \Biggl\{C^{\ell}_{0}
\ln^{2\ell-2}(a) + C^{\ell}_{1} \ln^{2\ell-3}(a) + \ldots + C^{\ell}_{2\ell-4}
\ln^2(a) \Biggr\} \; .
\end{equation}
The $\lambda^{\ell} C^{\ell}_{0} \ln^{2\ell-2}(a)$ terms are the {\it leading 
logarithms} at $\ell$ loop order; the remaining terms are {\it subdominant 
logarithms}. Assuming that the numerical coefficients $C^{\ell}_{k}$ are of 
order one, we see that the leading infrared logarithms all become order one
at $\ln(a) \sim 1/\sqrt{\lambda}$. At this time the highest subdominant
logarithm terms are still perturbatively small ($\sim \sqrt{\lambda}$), so 
it seems reasonable to attempt to follow the nonperturbative evolution by
resuming the series of leading infrared logarithms,
\begin{equation}
\rho_{\rm lead} = H^4 \sum_{\ell = 0}^{\infty} C^{\ell}_0 \Bigl(\lambda
\ln^2(a)\Bigr)^{\ell-1} \; .
\end{equation}
This is known as the {\it leading logarithm approximation}.

Starobinski\u{\i} has long maintained that his stochastic field equations
reproduce the leading logarithm approximation \cite{AAS}. With Yokoyama he
exploited this conjecture to explicitly solve for the nonperturbative,
late time limit of any model of the form \cite{SY},
\begin{equation}
\mathcal{L} = -\frac12 \partial_{\mu} \varphi \partial_{\nu} \varphi g^{\mu\nu}
\sqrt{-g} - V(\varphi) \sqrt{-g} \; , \label{potmod}
\end{equation}
assuming only that the potential $V(\varphi)$ is bounded below. When the 
potential is unbounded below the conjecture still gives the leading infrared
logarithms at each order, however, the theory fails to approach a static
limit.

Starobinski\u{\i}'s conjecture has recently been proved to all orders 
\cite{RPW2,TW4}. The field equations are first rewritten in Yang-Feldman form 
\cite{YF}, then the free field mode expansion is truncated at horizon
crossing, and the free field mode functions are replaced with their leading 
long wave length forms. This procedure converts the original quantum field
into a commuting random variable, but it preserves the leading infrared
logarithms. Although it was not evident at the time, the reason the field
can be infrared truncated is that every pair of fields in a simple potential 
model of the form (\ref{potmod}) is capable of inducing an infrared logarithm,
and the leading log term derives from requiring them to do so. Because only 
the infrared part of the field is responsible for infrared logarithms, we can
truncate and take the long wave length limit of the mode functions,
$u(t,k) \rightarrow H/\sqrt{2 k^3}$.\footnote{It is also necessary to take
the first nonzero term in the long wavelength limit for the retarded Green's
function of the Yang-Feldman equation. This turns out to require a higher 
order term in $u(t,k)$.} For example, the infrared logarithm 
in (\ref{lowest}) is,
\begin{equation}
\Bigl\langle \Omega \Bigl\vert \varphi^2(x) \Bigr\vert \Omega \Bigr\rangle_0 = 
\int \frac{d^3k}{(2\pi)^3} \, \Vert u(t,k)\Vert^2 \longrightarrow
\int_{H}^{H a} \frac{dk}{k} \frac{H^2}{4\pi^2} = 
\frac{H^2}{4 \pi^2} \ln(a) \; . 
\end{equation}

A field which can generate infrared logarithms is called {\it active}. Scalar
potential models of the form (\ref{potmod}) possess only active fields. 
However, more general theories can possess fields which are not themselves
capable of engendering an infrared logarithm. We call these {\it passive}
fields. A example of such a model is SQED,
\begin{equation}
\mathcal{L} = -\frac14 F_{\mu\nu} F_{\rho\sigma} g^{\mu\rho} g^{\nu\sigma}
\sqrt{-g} - \Bigl(\partial_{\mu} \!-\! i e A_{\mu}\Bigr) \varphi^*
\Bigl(\partial_{\nu} \!+\! i e A_{\nu}\Bigr) \varphi g^{\mu\nu} \sqrt{-g} \; .
\end{equation}
In this model the charged scalar is active whereas the photon is passive.

\begin{center}
\begin{picture}(300,100)(0,0)
\PhotonArc(150,60)(20,0,360){3}{15}
\DashArrowArc(150,30)(10,0,360){3}
\Vertex(150,80){3}
\Text(150,85)[b]{$x$}
\Vertex(150,40){3}
\Text(150,45)[b]{$x'$}
\end{picture}
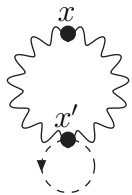
\\ {\rm Fig.~1: Two loop contribution to $\langle \Omega \vert F_{\mu\nu}(x) 
F_{\rho\sigma}(x)\vert \Omega \rangle$.}
\end{center}

Although passive fields cannot cause infrared logarithms, they can propagate
their effects. That is, an expectation value of passive fields can acquire 
an infrared logarithm from a loop correction involving an active field.
For example, the diagram in Fig.~1 gives a contribution to $\langle \Omega 
\vert F_{\mu\nu}(x) F_{\rho\sigma}(x)\vert \Omega \rangle$ which acquires 
an infrared logarithm through the scalar loop at the bottom.

Passive fields can also induce interactions between active fields. For 
example, the photon loop in Fig.~2 induces an effective $(\varphi^* \varphi)^2$
interaction in SQED,
\begin{center}
\begin{picture}(300,100)(0,0)
\PhotonArc(150,50)(30,0,180){3}{10}
\PhotonArc(150,50)(30,180,360){3}{10}
\DashArrowLine(120,50)(90,80){3}
\DashArrowLine(90,30)(120,50){3}
\DashArrowLine(180,50)(210,30){3}
\DashArrowLine(210,80)(180,50){3}
\Vertex(120,50){3}
\Vertex(180,50){3}
\end{picture}
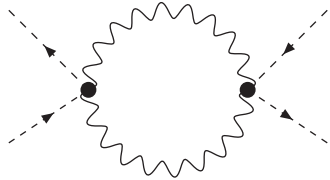
\\ {\rm Fig.~2: Effective $(\varphi^* \varphi)^2$ coupling in SQED.}
\end{center}

SQED --- and also gravity --- feature another complication in which
derivatives of active fields can induce interactions between undifferentiated
active fields. For example, the 3-point interaction of SQED,
\begin{equation}
i e \Bigl( \varphi^* \partial_{\mu} \varphi - \partial_{\mu} \varphi^* 
\varphi\Bigr) A_{\nu} g^{\mu\nu} \sqrt{-g} \; ,
\end{equation}
can induce an effective $\varphi^* \varphi$ coupling through the diagram
of Fig.~3,
\begin{center}
\begin{picture}(300,70)(0,0)
\PhotonArc(150,20)(30,0,180){3}{10}
\DashArrowLine(240,20)(180,20){3}
\DashArrowLine(180,20)(120,20){3}
\DashArrowLine(120,20)(60,20){3}
\Vertex(120,20){3}
\Text(120,5)[b]{$\varphi^* \partial \varphi$}
\Vertex(180,20){3}
\Text(180,5)[b]{$\partial \varphi^* \varphi$}
\end{picture}
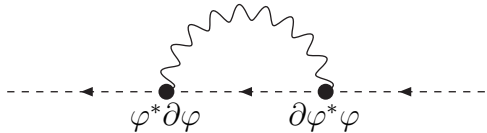
\\ {\rm Fig.~3: Effective $\varphi^* \varphi$ coupling in SQED.}
\end{center}
This is part of the full 1PI 2-point function which has recently been 
computed at one loop order \cite{KW}.

In generalizing Starobinski\u{\i}'s technique to theories which include
passive fields, and/or differentiated active fields, it is crucial to
realize that {\it the ultraviolet parts of passive fields and differentiated
active fields contribute on an equal footing with the infrared parts} in 
propagating infrared logarithms and in mediating interactions between 
undifferentiated active fields. So one cannot infrared truncate the passive 
fields, or even differentiated active fields. Instead the correct procedure 
is:
\begin{enumerate}
\item{Integrate out the passive fields and renormalize the resulting
effective action.}
\item{Integrate out the differentiated active fields and renormalize the 
resulting effective action. Note that this can always be done because the
original action is at most quadratic in derivatives.}
\item{Infrared truncate and stochastically simplify the effective action
of active fields.}
\end{enumerate}
One might suspect that the third step is not possible owing to the nonlocality
of the effective action. However, this nonlocality must be mediated by 
differential operators which, {\it precisely because they derive from inactive
fields}, must contain positive powers of the scale factor whose rapid time 
dependence weights the integral overwhelmingly at its upper limit and totally 
dominates the logarithms which might derive from the active fields. A typical 
example is the integral,
\begin{equation}
\int_0^t dt' {a'}^2 \ln(a') = \int_0^t dt' e^{2 H t'} H t' = \frac{{a'}^2}{2 H}
\Bigl\{ \ln(a') - \frac12\Bigr\} \Biggr\vert_0^t \simeq \frac1{2H} a^2 \ln(a) 
\; .
\end{equation}
For $\ln(a) \gg 1$ it is as though we simply divide the integrand by $2H$ and
evaluate it at the upper limit! Hence the hopelessly complicated ``effective 
action'' degenerates, in the leading log approximation, to a very tractable 
``effective potential,'' and the resulting local theory assumes the form 
(\ref{potmod}) already solved by Starobinski\u{\i} \cite{AAS,SY}.

Yukawa theory is especially simple because it possesses no differentiated
active fields, and because it is free of the subtle gauge fixing problems of
SQED \cite{KW} and gravity \cite{TW5,MW1}. In section 2 we review the full 
apparatus of perturbation theory for massless Yukawa on a locally de Sitter 
background. In section 3 we integrate out the fermion and renormalize the 
effective potential. A
curious and possibly significant property of Yukawa theory is that its
effective potential is unbounded below, a fact that survives in the flat
space limit and has even earned a place in Standard Model parameter
estimation. We check the stochastic formalism against an explicit two loop 
vacuum expectation value in section 4. In section 5 we employ the stochastic
formalism to obtain the leading log approximation for the Yukawa stress
tensor. Because the effective potential depends upon the inflationary Hubble
constant, the induced vacuum energy of the stress tensor does not quite agree 
with it. In fact the latter is initially positive whereas the former is always
negative. However, their asymptotic large field behaviors are identical, so
there seems no avoiding the conclusion that gravitational back-reaction in
this model must eventually halt inflation, although not in an acceptable
fashion. Our conclusions comprise section 6. An appendix presents the less
interesting, technical analysis behind the perturbative calculation described
in section 4.

\section{Massless Yukawa in de Sitter}

The coupling of gravity to particles with half integer spin is usually
accomplished by shifting the fundamental gravitational field variable 
from the metric $g_{\mu\nu}(x)$ to the vierbein $e_{\mu m}(x)$, although
there are other approaches \cite{HAW}. Greek letters stand for coordinate 
indices, Latin letters denote Lorentz indices, and both sorts take values 
in the set $\{0,1,2,\dots,(D\!-\!1)\}$. One recovers the metric by 
contracting two vierbeins into the Lorentz metric $\eta^{bc}$,
\begin{equation}
g_{\mu\nu}(x) = e_{\mu b}(x) e_{\nu c}(x) \eta^{bc} \; .
\end{equation}
The coordinate index is raised and lowered with the metric ($e^{\mu}\,_{b} 
= g^{\mu\nu} e_{\nu b}$), while the Lorentz index is raised and lowered with 
the Lorentz metric ($e_{\mu}\,^{b} = \eta^{bc} e_{\mu c}$). We employ the
usual metric-compatible and vierbein compatible connections,
\begin{eqnarray}
g_{\rho\sigma ; \mu} = 0 & \Longrightarrow & \Gamma^{\rho}_{~\mu\nu} =
\frac12 g^{\rho\sigma} \Bigl(g_{\sigma \mu , \nu} + g_{\nu \sigma , \mu}
- g_{\mu\nu , \sigma}\Bigr) \; , \\
e_{\beta b ; \mu} = 0 & \Longrightarrow & A_{\mu c d} = e^{\nu}_{~c} \Bigl(
e_{\nu d, \mu} - \Gamma^{\rho}_{~\mu\nu} e_{\rho d}\Bigr) \; . \label{spin}
\end{eqnarray}

Fermions also require gamma matrices, $\gamma^b_{ij}$. We assume their spinor
indices run over $i\!=\! 1, \ldots ,4$ in any dimension. The anti-commutation
relations are,
\begin{equation}
\Bigl\{\gamma^b , \gamma^c\Bigr\} \equiv \Bigl(\gamma^b \gamma^c + 
\gamma^c \gamma^b\Bigr) = -2 \eta^{bc} I \; .
\end{equation}
The Dirac Lorentz representation matrices are,
\begin{equation}
J^{bc} \equiv \frac{i}4 \Bigl[\gamma^b , \gamma^c\Bigr] = \frac{i}4 
\Bigl(\gamma^b \gamma^c - \gamma^c \gamma^b\Bigr) \; .
\end{equation}
They can be combined with the spin connection (\ref{spin}) to form the Dirac
covariant derivative operator,
\begin{equation}
\mathcal{D}_{\mu} \equiv \partial_{\mu} + \frac{i}2 A_{\mu cd} J^{cd} \; .
\end{equation}

In a general vierbein background the bare Lagrangian of massless, minimally 
coupled, Yukawa scalars with massless fermions is,
\begin{eqnarray}
\lefteqn{\mathcal{L} = -\frac12 \partial_{\alpha} \varphi
\partial_{\beta} \varphi g^{\alpha\beta} \sqrt{-g} - \frac12 \xi_0 
\varphi^2 R \sqrt{-g} - \frac1{4!} \lambda_0 \varphi^4 
\sqrt{-g} } \nonumber \\
& & \hspace{4cm} + i \overline{\psi} e^{\beta}_{~b} \gamma^b 
\mathcal{D}_{\beta} \psi \sqrt{-g} - f_0 \varphi
\overline{\psi} \psi \sqrt{-g} \; . \qquad \label{bareL}
\end{eqnarray}
The bare fields in this expression are $\varphi(x)$, $\overline{\psi}_i(x)$
and $\psi_i(x)$. The symbols $\xi_0$, $\lambda_0$ and
$f_0$ stand for, respectively, the bare conformal coupling, the bare 4-point 
coupling constant and the bare Yukawa coupling constant. Neither the scalar nor
the fermion requires a mass term because we desire the special model with zero 
renormalized masses, and no mass counterterms are required because mass is 
multiplicatively renormalized in dimensional regularization.

Renormalization is begun by expressing the bare fields in terms of the
renormalized ones,
\begin{equation}
\varphi \equiv \sqrt{Z} \, \varphi_{\rm r} \quad , \quad \overline{\psi}
\equiv \sqrt{Z_2} \, \overline{\psi}_{\rm r} \quad {\rm and} \quad \psi
\equiv \sqrt{Z_2} \, \psi_{\rm r} \; . \label{Zs}
\end{equation}
Substituting (\ref{Zs}) into the bare Lagrangian (\ref{bareL}) gives,
\begin{eqnarray}
\lefteqn{\mathcal{L} = -\frac12 Z \partial_{\alpha} \varphi_{\rm r}
\partial_{\beta} \varphi_{\rm r} g^{\alpha\beta} \sqrt{-g} - \frac12 Z \xi_0 
\varphi_{\rm r}^2 R \sqrt{-g} - \frac1{4!} Z^2 \lambda_0 \varphi_{\rm r}^4 
\sqrt{-g} } \nonumber \\
& & \hspace{3.5cm} + i Z_2 \overline{\psi}_{\rm r} e^{\beta}_{~b} \gamma^b
\mathcal{D}_{\beta} \psi_{\rm r} \sqrt{-g} - \sqrt{Z} Z_2 f_0 \varphi_{\rm r}
\overline{\psi}_{\rm r} \psi_{\rm r} \sqrt{-g} \; . \qquad \label{L1}
\end{eqnarray}
We next enforce the conditions that the renormalized scalar should have neither
a conformal coupling nor a 4-point coupling,
\begin{equation}
Z \xi_0 \equiv 0 + \delta \xi_0 \quad , \quad Z^2 \lambda_0 \equiv 0 + \delta
\lambda_0 \quad {\rm and} \quad \sqrt{Z} Z_2 f_0 = f + \delta\!f \; .
\end{equation}
Of course the model {\it does} require conformal and 4-point counterterms! We 
also define the field strengths as usual,
\begin{equation}
Z \equiv 1 + \delta Z \qquad {\rm and} \qquad Z_2 \equiv 1 + \delta Z_2 \; .
\end{equation}
The structure of renormalized perturbation theory is complete when we express
the Lagrangian in terms of primitive interactions and counterterms,
\begin{eqnarray}
\lefteqn{\mathcal{L} = -\frac12 \partial_{\alpha} \varphi_{\rm r} 
\partial_{\beta} \varphi_{\rm r} g^{\alpha\beta} \sqrt{-g} + i \overline{
\psi}_{\rm r} e^{\beta}_{~b} \gamma^b \mathcal{D}_{\beta} \psi_{\rm r}
\sqrt{-g} - f \varphi_{\rm r} \overline{\psi}_{\rm r} \psi_{\rm r} \sqrt{-g}
} \nonumber \\
& & \hspace{2cm} -\frac{\delta Z}2 \partial_{\alpha} \varphi_{\rm r}
\partial_{\beta} \varphi_{\rm r} g^{\alpha\beta} \sqrt{-g} + i \delta Z_2 
\overline{\psi}_{\rm r} e^{\beta}_{~b} \gamma^b \mathcal{D}_{\beta} \psi_{\rm r}
\sqrt{-g} \nonumber \\
& & \hspace{4cm} - \frac{\delta \xi}2 \varphi_{\rm r}^2 R \sqrt{-g} - 
\frac{\delta \lambda}{4!} \varphi_{\rm r}^4 \sqrt{-g} - \delta\!f 
\varphi_{\rm r} \overline{\psi}_{\rm r} \psi_{\rm r} \sqrt{-g} \; . \qquad 
\label{L2}
\end{eqnarray}

The preceding analysis has so far been for a general geometry. Of course we
are interested in the special case of inflation, for which de Sitter is an
excellent paradigm. We work on the open submanifold of $D$ dimensional de 
Sitter space in the conformal coordinate system for which the invariant 
element is,
\begin{equation}
ds^2 = a^2(\eta) \Bigl(-d\eta^2 + d\vec{x} \!\cdot\! d\vec{x}\Bigr) \qquad
{\rm and} \qquad a(\eta) = -\frac1{H \eta} \; .
\end{equation}
Of course this makes the metric $g_{\mu\nu} = a^2(\eta) \eta_{\mu\nu}$. We 
also employ the ghost-free, Lorentz symmetric gauge \cite{RPW3} which 
determines the vierbein,
\begin{equation}
e_{\mu b} = e_{b \mu} \qquad \Longrightarrow \qquad e_{\mu b} = a \eta_{\mu b}
\; .
\end{equation}
At this stage there is no more point in distinguishing Lorentz indices from
coordinate ones. With these conventions the covariant derivative operator 
takes the simple form,
\begin{equation}
\mathcal{D}_{\mu} \longrightarrow \partial_{\mu} + \frac14 H a \Bigl[\gamma^0,
\gamma_{\mu}\Bigr] \; .
\end{equation}
The special case of its contraction into $e^{\mu}_{~b} \gamma^b$ is even
simpler,
\begin{equation}
\gamma^b e^{\mu}_{~b} \mathcal{D}_{\mu} \equiv \; \not{\hspace{-.15cm} 
\mathcal{D}} \longrightarrow a^{-(\frac{D+1}2)} \hspace{-.1cm} \not{
\hspace{-.1cm} \partial} a^{(\frac{D-1}2)} \equiv a^{-(\frac{D+1}2)} 
\gamma^{\mu} \partial_{\mu} a^{(\frac{D-1}2)} \; .
\end{equation}

The scalar and fermion propagators can be largely expressed in terms of the 
following function of the invariant length $\ell(x;x')$ between $x^{\mu}$ and 
$x^{\prime \mu}$,
\begin{eqnarray}
y(x;x') & \equiv & 4 \sin^2\Bigl(\frac12 H \ell(x;x')\Bigr) 
= a a' H^2 {\Delta x }^2(x;x') \; , \\
& = & a a' H^2 \Bigl( \Vert \vec{x} - \vec{x}'\Vert^2 - (\vert \eta \!-\! 
\eta'\vert \!-\! i\delta)^2 \Bigr) \; . \label{fully}
\end{eqnarray}
The most singular term for each case involves the propagator for a massless,
conformally coupled scalar,
\begin{equation}
{i\Delta}_{\rm cf}(x;x') = \frac{H^{D-2}}{(4\pi)^{\frac{D}2}} \Gamma\Bigl(
\frac{D}2 \!-\! 1\Bigr) \Bigl(\frac4{y}\Bigr)^{\frac{D}2-1} \; .
\end{equation}
It has long been known that there is no de Sitter invariant solution for 
the propagator of a massless, minimally coupled scalar \cite{AF}. If one 
elects to break de Sitter invariance while preserving homogeneity and 
isotropy --- this is known as the ``E(3)'' vacuum \cite{BA} --- the minimal 
solution is \cite{OW1,OW2},
\begin{eqnarray}
\lefteqn{i \Delta(x;x') =  i \Delta_{\rm cf}(x;x') } \nonumber \\
& & + \frac{H^{D-2}}{(4\pi)^{\frac{D}2}} \frac{\Gamma(D \!-\! 1)}{\Gamma(
\frac{D}2)} \left\{\! \frac{D}{D\!-\! 4} \frac{\Gamma^2(\frac{D}2)}{\Gamma(D
\!-\! 1)} \Bigl(\frac4{y}\Bigr)^{\frac{D}2 -2} \!\!\!\!\!\! - \pi 
\cot\Bigl(\frac{\pi}2 D\Bigr) + \ln(a a') \!\right\} \nonumber \\
& & + \frac{H^{D-2}}{(4\pi)^{\frac{D}2}} \! \sum_{n=1}^{\infty}\! \left\{\!
\frac1{n} \frac{\Gamma(n \!+\! D \!-\! 1)}{\Gamma(n \!+\! \frac{D}2)} 
\Bigl(\frac{y}4 \Bigr)^n \!\!\!\! - \frac1{n \!-\! \frac{D}2 \!+\! 2} 
\frac{\Gamma(n \!+\!  \frac{D}2 \!+\! 1)}{\Gamma(n \!+\! 2)} \Bigl(\frac{y}4
\Bigr)^{n - \frac{D}2 +2} \!\right\} \! . \quad \label{Delta}
\end{eqnarray}
This expression may seem daunting but it is actually simple to use because
the infinite sum vanishes in $D\!=\!4$, and the terms of this sum goes like
higher and higher powers of $y(x;x')$. Hence the infinite sum can only 
contribute when multiplied by a divergent term, and even then only the first
few terms can contribute.

It is useful, in the stochastic analysis to follow, for us to consider the
fermion propagator in the presence of an arbitrary, potentially spacetime
dependent, scalar field $\varphi(x)$,
\begin{equation}
i \mathcal{S}[f \varphi](x;x') \equiv \Bigl\langle x \Bigl\vert \frac{i}{\sqrt{
-g} ( i \hspace{-.08cm} \not{\hspace{-.15cm} \mathcal{D}} \, \!-\! f \varphi)}
\Bigr\vert x' \Bigr\rangle \; .
\end{equation}
Of course this can only be evaluated for a handful of field configurations.
The case of a constant, $f \varphi(x) \!=\! m$, was solved by Candelas 
and Raine \cite{CR},
\begin{eqnarray}
\lefteqn{i \mathcal{S}[m](x;x') = \frac{H^{D-2}}{(4\pi)^{\frac{D}2}} 
\Gamma\Bigl(\frac{D}2 \!-\! 1\Bigr) \Bigl(a i \hspace{-.08cm} \not{\hspace{
-.15cm} \mathcal{D}} \; \frac1{\sqrt{a a'}} \!+\! \sqrt{\frac{a}{a'}} m \, 
I \Bigr) } \nonumber \\
& & \hspace{-.7cm} \times \Biggl\{ \frac{\Gamma(\frac{D}2 \!-\! 1 \!+\! i 
\frac{m}{H}) \Gamma(\frac{D}2 \!-\! i \frac{m}{H})}{\Gamma(\frac{D}2 \!-\! 1) 
\Gamma(\frac{D}2)} \mbox{}_2 F_1\Bigl(\frac{D}2 \!-\! 1 \!+\! i \frac{m}{H},
\frac{D}2 \!-\! i \frac{m}{H}; \frac{D}2 ;1 \!-\! \frac{y}4\Bigr) \Bigl(\frac{
I \!-\! \gamma^0}{2}\Bigr) \nonumber \\
& & \hspace{-.7cm} + \frac{\Gamma(\frac{D}2 \!-\! 1 \!-\! i 
\frac{m}{H}) \Gamma(\frac{D}2 \!+\! i \frac{m}{H})}{\Gamma(\frac{D}2 \!-\! 1) 
\Gamma(\frac{D}2)} \mbox{}_2 F_1\Bigl(\frac{D}2 \!-\! 1 \!-\! i \frac{m}{H},
\frac{D}2 \!+\! i \frac{m}{H}; \frac{D}2 ;1 \!-\! \frac{y}4\Bigr) \Bigl(\frac{
I \!+\! \gamma^0}{2}\Bigr) \Biggr\} . \qquad \label{fullp}
\end{eqnarray}
The propagator we use in perturbative calculations is the massless limit of
this one,
\begin{eqnarray}
i\Bigl[S\Bigr](x;x') & = & \frac{H^{D-2}}{(4\pi)^{\frac{D}2}} \Gamma\Bigl(
\frac{D}2 \!-\! 1\Bigr) a i \hspace{-.08cm} \not{\hspace{-.15cm} 
\mathcal{D}} \; \frac1{\sqrt{a a'}} \mbox{}_2 F_1\Bigl(\frac{D}2 \!-\! 1 ,
\frac{D}2 ; \frac{D}2 ;1 \!-\! \frac{y}4\Bigr) \; , \qquad \\
& = & a i \hspace{-.08cm} \not{\hspace{-.15cm} \mathcal{D}} \; \frac1{
\sqrt{a a'}} \, i\Delta_{\rm cf}(x;x') \; , \\
& = & (a a')^{-(\frac{D-1}2)} \times i \hspace{-.08cm} \not{\hspace{-.10cm} 
\partial} \; \times \frac{\Gamma(\frac{D}2 \!-\! 1)}{4 \pi^{\frac{D}2}}
\frac1{\Delta x^{D-2}} \; . \label{fermp}
\end{eqnarray}
The final expression is just a conformal rescaling of the propagator for a
massless fermion in flat space, as it should be in conformal coordinates.

It is useful to recast the Candelas-Raine solution (\ref{fullp}) using the 
transformation formula of hypergeometric functions (see expression 9.131.2 
in \cite{GR}),
\begin{eqnarray}
\lefteqn{i \mathcal{S}[m](x;x') = \frac{H^{D-2}}{(4\pi)^{\frac{D}2}} 
\Gamma\Bigl(\frac{D}2 \!-\! 1\Bigr) \Bigl(a i \hspace{-.08cm} \not{\hspace{
-.15cm} \mathcal{D}} \; \frac1{\sqrt{a a'}} \!+\! \sqrt{\frac{a}{a'}} m \, 
I \Bigr) } \nonumber \\
& & \hspace{-.5cm} \times \Biggl\{ \Biggl[ \frac{\Gamma(1 \!-\! \frac{D}2)
\Gamma(\frac{D}2 \!-\! 1 \!+\! i \frac{m}{H}) \Gamma(\frac{D}2 \!-\! i 
\frac{m}{H})}{\Gamma(\frac{D}2 \!-\! 1) \Gamma(i\frac{m}{H}) \Gamma(1 \!-\!
i\frac{m}{H})} \mbox{}_2 F_1\Bigl(\frac{D}2 \!-\! 1 \!+\! i \frac{m}{H},
\frac{D}2 \!-\! i \frac{m}{H}; \frac{D}2 ;\frac{y}4\Bigr) \nonumber \\
& & \hspace{3cm} + \Bigl(\frac{y}4\Bigr)^{1-\frac{D}2} \mbox{}_2 F_1\Bigl(1 
\!-\! i \frac{m}{H},i \frac{m}{H};2 \!-\! \frac{D}2 ;\frac{y}4\Bigr) \Biggr]
\Bigl(\frac{I \!-\! \gamma^0}{2}\Bigr) \nonumber \\
& & + \Biggl[ \frac{\Gamma(1 \!-\! \frac{D}2)
\Gamma(\frac{D}2 \!-\! 1 \!-\! i \frac{m}{H}) \Gamma(\frac{D}2 \!+\! i 
\frac{m}{H})}{\Gamma(\frac{D}2 \!-\! 1) \Gamma(-i\frac{m}{H}) \Gamma(1 \!+\!
i \frac{m}{H})} \mbox{}_2 F_1\Bigl(\frac{D}2 \!-\! 1 \!-\! i \frac{m}{H},
\frac{D}2 \!+\! i \frac{m}{H}; \frac{D}2 ;\frac{y}4\Bigr) \nonumber \\
& & \hspace{3cm} + \Bigl(\frac{y}4\Bigr)^{1-\frac{D}2} \mbox{}_2 F_1\Bigl(1 
\!+\! i \frac{m}{H},-i \frac{m}{H};2 \!-\! \frac{D}2 ;\frac{y}4\Bigr) \Biggr]
\Bigl(\frac{I \!+\! \gamma^0}{2}\Bigr) \Biggr\} . \qquad
\end{eqnarray}
In dimensional regularization all $D$-dependent powers of $y(x;x')$ vanish
at coincidence. Hence we obtain,\footnote{Note that the back-acting covariant 
derivative is $\overleftarrow{\mathcal{D}}_{\mu} = \overleftarrow{\partial}_{
\mu} - \frac14 H a [\gamma^0,\gamma_{\mu}]$.}
\begin{eqnarray}
\lim_{x' \rightarrow x} i\mathcal{S}[m](x;x') \!\!\!\!& = &\!\!\!\! \frac{m 
H^{D-2}}{(4\pi)^{\frac{D}2}} \frac{\Gamma(\frac{D}2 \!+\! i \frac{m}{H})\Gamma(
\frac{D}2 \!-\! i \frac{m}{H})\Gamma(1 \!-\! \frac{D}2)}{\Gamma(1 \!+\! i
\frac{m}{H}) \Gamma(1 \!-\! i\frac{m}{H})} \!\times\! I \; , \label{C1} \\
\lim_{x' \rightarrow x} \mathcal{D}_{\mu} i\mathcal{S}[m](x;x') \!\!\!\!& = & 
\!\!\!\! \frac{m H^{D-2}}{(4\pi)^{\frac{D}2}} \frac{\Gamma(\frac{D}2 \!+\! i 
\frac{m}{H}) \Gamma(\frac{D}2 \!-\! i \frac{m}{H}) \Gamma(1 \!-\! \frac{D}2)}{
\Gamma(1 \!+\! i\frac{m}{H}) \Gamma(1 \!-\! i\frac{m}{H})} \!\times\! 
\frac{i}{D} m a \gamma_{\mu} \; , \label{C2} \\
\lim_{x' \rightarrow x} i\mathcal{S}[m](x;x')\overleftarrow{\mathcal{D}}_{\mu}'
\!\!\!\! & = & \!\!\!\! \frac{m H^{D-2}}{(4\pi)^{\frac{D}2}} \frac{\Gamma(
\frac{D}2 \!+\! i \frac{m}{H}) \Gamma(\frac{D}2 \!-\! i \frac{m}{H}) \Gamma(1 
\!-\! \frac{D}2)}{\Gamma(1 \!+\! i \frac{m}{H}) \Gamma(1 \!-\! i\frac{m}{H})} \!
\times\! -\frac{i}{D} m a \gamma_{\mu} . \qquad \label{C3}
\end{eqnarray}

\section{Stochastic Effective Action}

Integrating out the fermions gives rise to a scalar effective action,
\begin{equation}
e^{i \Gamma[\varphi]} = \Fint [d\overline{\psi}] [d\psi] \, e^{i S[\varphi,
\overline{\psi},\psi]} =  e^{i S_{\rm s}[\varphi]} \det\Bigl[ \sqrt{-g} \Bigl(i 
\hspace{-.08cm} \not{\hspace{-.15cm} \mathcal{D}} \, \!-\! f_0 \varphi\Bigr)
\Bigr] \; .
\end{equation}
Here $S_{\rm s}[\varphi]$ stands for the action associated with the purely 
scalar part of the bare Lagrangian,
\begin{equation}
\mathcal{L}_{\rm s} = -\frac12 \partial_{\mu} \varphi \partial_{\nu} \varphi
g^{\mu\nu} \sqrt{-g} - \frac{\xi_0}2 \varphi^2 R \sqrt{-g} - \frac{\lambda_0}4
\varphi^4 \sqrt{-g} \; .
\end{equation}
The exact effective scalar field equation is,
\begin{eqnarray} 
\lefteqn{\frac{\delta \Gamma[\varphi]}{\delta \varphi(x)} = \partial_{\mu}
\Bigl(\sqrt{-g} g^{\mu\nu} \partial_{\nu} \varphi(x)\Bigr) - \xi_0
\varphi(x) R \sqrt{-g} - \frac{\lambda_0}6 \varphi^3(x) \sqrt{-g}} \nonumber \\
& & \hspace{3.5cm} - {\rm Tr}\Biggl[ \frac{i}{\sqrt{-g} (i \hspace{-.08cm} 
\not{\hspace{-.15cm} \mathcal{D}} \, \!-\! f_0 \varphi)} \frac{\delta}{\delta
\varphi(x)} \sqrt{-g} \Bigl(i \hspace{-.08cm} \not{\hspace{-.15cm} \mathcal{D}}
\, \!-\! f_0 \varphi\Bigr) \Biggr] \; , \\
& & = \partial_{\mu} \Bigl(\sqrt{-g} g^{\mu\nu} \partial_{\nu} \varphi(x)\Bigr)
- \xi_0 \varphi(x) R \sqrt{-g} - \frac{\lambda_0}6 \varphi^3(x) \sqrt{-g} 
\nonumber \\
& & \hspace{3.5cm} + f_0 {\rm Tr}\Bigl[i\mathcal{S}[f_0 \varphi](x;x)\Bigr] 
\sqrt{-g} \; . \qquad \label{exact}
\end{eqnarray}

Were we trying to solve the full quantum field theory, equation (\ref{exact})
would be a dead end because we lack an explicit expression for the coincidence 
limit of the fermion propagator in the presence of a general $\varphi(x)$.
However, we are focused instead on the leading infrared logarithms, and this
fact permits a crucial simplification: {\it we can evaluate the fermion 
propagator as if the scalar were constant!} Making use of (\ref{C1}) we
infer a local field equation which agrees exactly with the full theory in
the leading log approximation,
\begin{eqnarray} 
\lefteqn{\frac{\delta \Gamma[\varphi]}{\delta \varphi(x)} \longrightarrow
\partial_{\mu} \Bigl(\sqrt{-g} g^{\mu\nu} \partial_{\nu} \varphi\Bigr) - \xi_0
\varphi R \sqrt{-g} - \frac{\lambda_0}6 \varphi^3 \sqrt{-g} } \nonumber \\
& & \hspace{2.5cm} + \frac{4 f_0^2 H^{D-2}}{(4\pi)^{\frac{D}2}} \frac{\Gamma(
\frac{D}2 \!+\! i \frac{f_0 \varphi}{H}) \Gamma(\frac{D}2 \!-\! i \frac{f_0 
\varphi}{H}) \Gamma(1 \!-\! \frac{D}2)}{\Gamma(1 \!+\! i\frac{f_0 \varphi}{H}) 
\Gamma(1 \!-\! i\frac{f_0 \varphi}{H})} \varphi \sqrt{-g} \; . \qquad
\label{local}
\end{eqnarray}
Of course the final term in (\ref{local}) is also related to the effective
potential (it is $-V_{\rm eff}'(\varphi) \sqrt{-g}$) and has appeared many 
times before in this guise \cite{CR,IMM,IMO}.

The factor of $\Gamma(1 \!-\! \frac{D}2)$ in equation (\ref{local}) is the
only divergence we shall see in the stochastic formalism. It can be removed 
using the parameters of the scalar potential, $\xi_0$ and $\lambda_0$. In 
particular, the stochastic formalism does not require either field strength 
renormalization or renormalization of the Yukawa coupling,
\begin{equation}
\delta Z = \delta Z_2 = \delta\!f = 0 \qquad \Longrightarrow \qquad
\xi_0 = \delta \xi \quad {\rm and} \quad \lambda_0 = \delta \lambda \; .
\end{equation}
To renormalize (\ref{local}) we set $D \!=\! 4 \!-\! \epsilon$ and make use 
of the following expansions,
\begin{eqnarray}
\lefteqn{\Gamma\Bigl(1 \!-\! \frac{D}2\Bigr) = -\frac2{\epsilon} \!+\! \gamma 
\!-\!  1 + O(\epsilon) \; , } \\
\lefteqn{\frac{\Gamma(\frac{D}2 \!+\! i \frac{f \varphi}{H}) \Gamma(\frac{D}2 
\!-\! i \frac{f \varphi}{H})}{\Gamma(1 \!+\! i\frac{f \varphi}{H}) \Gamma(1 
\!-\! i \frac{f \varphi}{H})} = 1 \!+\! \Bigl(\frac{f\varphi}{H}\Bigr)^2 \!-\!
\epsilon } \nonumber \\
& & \hspace{2cm} - \Bigl[1 \!+\! \Bigl(\frac{f \varphi}{H}\Bigr)^2\Bigr]
\Biggl[\psi\Bigl(1 \!+\! i \frac{f \varphi}{H}\Bigr) \!+\! \psi\Bigl(1 \!-\! 
i \frac{f \varphi}{H}\Bigr)\Biggr] \frac{\epsilon}2 + O(\epsilon^2) \; . 
\qquad \label{gamrat}
\end{eqnarray}
The symbol ``$\psi(z)$'' in this last expression of course stands for the 
psi function rather than the fermi field (see section 8.36 of \cite{GR}),
\begin{equation}
\psi(1 \!+\! z) \equiv \frac{d}{dz} \ln\Bigl(\Gamma(1 \!+\!z)\Bigr) =
-\gamma + \sum_{n=2}^{\infty} (-1)^n \zeta(n) z^{n-1} \; .
\end{equation}
Note that (\ref{gamrat}) is both real and even in $\varphi$,
\begin{eqnarray}
\lefteqn{\frac{\Gamma(\frac{D}2 \!+\! i \frac{f \varphi}{H}) \Gamma(\frac{D}2 
\!-\! i \frac{f \varphi}{H})}{\Gamma(1 \!+\! i\frac{f \varphi}{H}) \Gamma(1 
\!-\! i \frac{f \varphi}{H})} = 1 \!+\! \Bigl(\frac{f\varphi}{H}\Bigr)^2 
\!-\! (1 \!-\! \gamma) \epsilon \!-\! \Bigl[\zeta(3) \!-\! \gamma\Bigr]
\Bigl(\frac{f \varphi}{H}\Bigr)^2 \epsilon } \nonumber \\
& & \hspace{3cm} - \sum_{n=2}^{\infty} (-1)^n \Bigl[\zeta(2n \!-\! 1) \!-\!
\zeta(2n\!+\!1)\Bigr] \Bigl(\frac{f \varphi}{H}\Bigr)^{2n} \epsilon
+ O(\epsilon^2) \; . \qquad
\end{eqnarray}
From these expansions it is apparent that we can renormalize so as to keep 
the scalar massless and free to order $f^6$,
\begin{eqnarray}
\delta \xi & = & \frac{4 f^2 H^{D-4}}{(4 \pi)^{\frac{D}2}} \frac{\Gamma(1 \!-\!
\frac{D}2)}{D (D \!-\! 1)} + \frac{f^2}{24 \pi^2} (1 \!-\! \gamma) \; , 
\label{dxi} \\
\delta \lambda & = & \frac{24 f^4 H^{D-4}}{(4 \pi)^{\frac{D}2}} \Gamma\Bigl(1 
\!-\! \frac{D}2\Bigr) + \frac{3 f^4}{\pi^2} \Bigl[ \zeta(3) \!-\! \gamma\Bigr] 
\; .
\end{eqnarray}
Substituting in (\ref{local}) and taking the limit $D \rightarrow 4$ gives
the following effective equation of motion,
\begin{eqnarray}
\lefteqn{\frac{\delta \Gamma[\varphi]}{\delta \varphi(x)} \longrightarrow
\partial_{\mu} \Bigl(\sqrt{-g} g^{\mu\nu} \partial_{\nu} \varphi\Bigr) }
\nonumber \\
& & \hspace{2cm} 
+ \frac{f H^3}{2 \pi^2} \sum_{n=2}^{\infty} (-1)^n \Bigl[ \zeta(2n \!-\! 1)
- \zeta(2n \!+\! 1)\Bigr] \Bigl(\frac{f \varphi}{H}\Bigr)^{2n+1} 
\sqrt{-g} \; , \qquad \\
& & = \partial_{\mu} \Bigl(\sqrt{-g} g^{\mu\nu} \partial_{\nu} \varphi\Bigr)
- \frac{f H^3}{2 \pi^2} \zeta(3) \Bigl(\frac{f \varphi}{H}\Bigr)^3 \sqrt{-g}
+ \frac{f H^3}{2 \pi^2} \Bigl[1 \!+\! \Bigl(\frac{f \varphi}{H}\Bigr)^2 \Bigr] 
\nonumber \\
& & \hspace{3cm} \times \Biggl[\gamma + \frac12 \psi\Bigl(1 \!+\! i \frac{f
\varphi}{H}\Bigr) + \frac12 \psi\Bigl(1 \!-\! i \frac{f \varphi}{H}\Bigr) 
\Biggr] \Bigl(\frac{f \varphi}{H}\Bigr) \sqrt{-g} \; . \qquad
\end{eqnarray}

It is well to digress at this point to establish an important correspondence
limit that bears upon the validity and physical interpretation of our 
renormalization condition (\ref{dxi}). Duffy and Woodard computed the one 
loop scalar self-mass-squared in this theory and used it to solve the 
effective scalar field equation \cite{DW}. They found (equations (38) and (77) 
in \cite{DW}) that the following choice for the conformal counterterm results 
in there being no significant late time corrections to the scalar mode 
functions at one loop order,
\begin{eqnarray}
\delta \xi_{\rm DW} & = & \frac{f^2 H^{D-4}}{(4 \pi)^{\frac{D}2}} \frac{(D\!-\!
2) \Gamma(\frac{D}2 \!-\!2)}{2 (D \!-\! 1) (D \!-\! 3)} + \frac{f^2}{32 \pi^2}
\; , \\
& = & \frac{f^2 H^{D-4}}{(4 \pi)^{\frac{D}2}} \Bigl(-\frac2{3 \epsilon}\Bigr)
- \frac{f^2}{16 \pi^2} \Bigl(\frac{\gamma}3 \!+\! \frac1{18}\Bigr) + O(
\epsilon) \; .
\end{eqnarray}
Up to irrelevant terms of order $\epsilon$, this is precisely the same 
renormalization (\ref{dxi}) as we have used! Therefore our stochastic 
renormalization conventions agree with the full theory in the regime of 
significant late time effects, just as they should.

Up to some different renormalization conventions, our expression for the
scalar effective potential agrees with equation (40) of Candelas and Raine
\cite{CR},
\begin{eqnarray}
V(\varphi) & = & -\frac{H^4}{4 \pi^2} \sum_{n=2}^{\infty} \frac{(-1)^n}{n \!+\!
1} \Bigl[ \zeta(2n \!-\! 1) - \zeta(2n \!+\! 1)\Bigr] \Bigl(\frac{f \varphi}{H}
\Bigr)^{2n+2} \; , \\
& = & -\frac{H^4}{8 \pi^2} \Biggl\{2 \gamma \Bigl(\frac{f\varphi}{H}\Bigr)^2
- [\zeta(3) \!-\! \gamma] \Bigl(\frac{f\varphi}{H}\Bigr)^4 \nonumber \\
& & \hspace{3.5cm} + 2 \int_0^{\frac{f\varphi}{H}} \!\!\!\!\! dx \, (x \!+\! 
x^3) \Bigl[\psi(1 \!+\! i x) \!+\! \psi(1 \!-\! i x)\Bigr] \Biggr\} . \qquad
\label{psiform}
\end{eqnarray}
We have now reduced the theory to a completely finite, scalar model of the
form already solved by Starobinski\u{\i} \cite{AAS,SY}. An interesting 
and possibly significant result of applying his technique is that {\it this 
model fails to approach a static limit at late times!} This is obvious once
one recognizes that the potential $V(\varphi)$ is unbounded from below.

One might expect that $V(\varphi)$ is negative because a nonzero scalar (of 
either sign) drives the fermion mass positive \cite{PW3,GP}, which must 
lower the vacuum energy. The absence of a lower bound is most easily proved 
by making use of the asymptotic expansion Stirling's formula implies for 
the psi function,
\begin{equation}
\psi(z) = \ln(z) - \frac1{2 z} - \frac1{12 z^2} + O\Bigl(\frac1{z^4}\Bigr) \; ,
\end{equation}
to evaluate the strong field limit of (\ref{psiform}),
\begin{eqnarray}
\lefteqn{V(\varphi) = -\frac{H^4}{8 \pi^2} \Biggl\{ \Bigl(\frac{f \varphi}{H}
\Bigr)^4 \ln\Bigl(\frac{f \vert \varphi\vert}{H}\Bigr) - \Bigl[\zeta(3) \!+\! 
\frac14 \!-\! \gamma\Bigr] \Bigl(\frac{f \varphi}{H} \Bigr)^4 } \nonumber \\
& & \hspace{1.5cm} + 2 \Bigl(\frac{f \varphi}{H}\Bigr)^2 \ln\Bigl(\frac{f \vert
\varphi\vert}{H}\Bigr) - \Bigl(\frac56 \!-\! 2 \gamma\Bigr) \Bigl(\frac{f 
\varphi}{H}\Bigr)^2 + O\Bigl(\ln\Bigl[\frac{f \vert \varphi\vert}{H}\Bigr]
\Bigr) \Biggr\} .
\end{eqnarray}
Of course the large $\varphi$ regime is also the small $H$ regime, at which
point we make contact with equation (6.10) in the classic paper by Coleman
and Weinberg \cite{CW}. The negative potential has long been recognized to
render pure Yukawa unstable in flat space \cite{SHR}. A constraint on the
Higgs mass can be derived in the Standard Model from the need to avoid this
instability for the large Yukawa coupling of the top quark \cite{FJJ}.

Although we have just seen that the instability is present in flat space, 
inflation does have a role to play. A flat space scalar would simply roll
down an unbounded, negative potential. However, the Hubble friction of 
expansion retards the scalar's downward progress. The potential's curvature
is slight, for small $f$, so the scalar's evolution is for a very long time
driven by the pressure of inflationary particle production. Only when the
scalar's magnitude approaches the nonperturbatively large scale of $\varphi
\sim H/f$ does the unbounded potential begin to dominate.

\section{Reality Check}

\begin{center}
\begin{picture}(300,100)(0,0)
\DashLine(150,85)(150,15){4}
\ArrowArc(150,50)(35,-90,90)
\ArrowArc(150,50)(35,90,270)
\Vertex(150,85){3}
\Text(150,93)[b]{$x$}
\Vertex(150,15){3}
\Text(150,0)[b]{$x'$}
\end{picture}
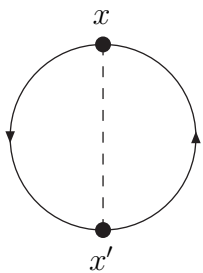
\\ {\rm Fig.~4: Lowest order contribution to $\langle \Omega \vert T[\varphi_{
\rm r}(x) \overline{\psi}_{\rm r}(x) \psi_{\rm r}(x)] \vert \Omega \rangle$.}
\end{center}

In this section we will test the stochastic formalism by comparing its
prediction with an explicit two loop evaluation of the coincident vertex 
function (see Fig.~4),
\begin{equation}
\Bigl\langle \Omega \Bigl\vert T\Bigl[\varphi_{\rm r}(x) \overline{\psi}_{\rm 
r}(x) \psi_{\rm r}(x)\Bigr] \Bigr\vert \Omega \Bigr\rangle \; . \label{test}
\end{equation}
Because there is no field strength renormalization in the stochastic
formalism, we can ignore the distinction between renormalized and 
unrenormalized fields in working out the stochastic prediction for 
(\ref{test}).
Integrating the fermions out exactly gives the trace of the field-dependent
fermion propagator, which we can again evaluate for constant field 
configurations in the leading logarithm approximation,
\begin{eqnarray}
\lefteqn{e^{-i \Gamma[\varphi]} \Fint [d\overline{\psi}] [d\psi] e^{iS[\varphi,
\overline{\psi},\psi]} \times \varphi(x) \overline{\psi}(x) \psi(x) =
\varphi(x) {\rm Tr}\Biggl[-i \mathcal{S}[f \varphi](x;x)\Biggr] \; ,} \\
& & \hspace{4cm} \longrightarrow -4 f \varphi^2(x) \frac{H^{D-2}}{(4 \pi)^{
\frac{D}2}} \Gamma\Bigl(1 \!-\! \frac{D}2\Bigr) \Biggl\Vert \frac{\Gamma(
\frac{D}2 \!+\! i \frac{f \varphi}{H})}{\Gamma(1 \!+\! i \frac{f \varphi}{H})} 
\Biggr\Vert^2 , \qquad \label{all} \\
& & \hspace{4cm} = -\frac{4 f H^{D-2}}{(4\pi)^{\frac{D}2}} \Gamma\Bigl(1 \!-\! 
\frac{D}2\Bigr) \Gamma^2\Bigl(\frac{D}2\Bigr) \varphi^2(x) + O(f^3) \; . 
\qquad \label{first}
\end{eqnarray}
Hence our prediction for the order $f$ infrared logarithm is divergent,
\begin{equation}
\Bigl\langle \Omega \Bigl\vert T\Bigl[\varphi_{\rm r}(x) \overline{\psi}_{\rm 
r}(x) \psi_{\rm r}(x)\Bigr] \Bigr\vert \Omega \Bigr\rangle \longrightarrow +
\frac{f H^4}{8 \pi^4} \frac{\ln(a)}{\epsilon} + {\rm finite} + O(f^3) \; .
\label{stochpred}
\end{equation}
Note that the divergence arises from integrating out coincident fermion fields.
It has nothing to do with the stochastic formalism {\it per se}, except for
being the correct leading log result for the expectation value we have chosen 
to compute.

That was easy! A measure of the power of the stochastic formalism is that we
could just as simply have obtained the leading log result at order $f^3$ or 
higher. We turn now to the much more difficult task of perturbatively computing
the full order $f$ result for comparison.\footnote{This computation was done in 
collaboration with P. M. Ho.} In models for which the ``in'' and ``out'' 
vacua differ, the in-out matrix elements computed with the usual Feynman 
rules are not true expectation values. To obtain an expectation value
such as (\ref{test}) one must employ the Schwinger-Keldysh formalism 
\cite{JS,M,BM,K,CSHY,J,CH}. For a recent review of the position-space
formalism see \cite{FW}. Here we simply summarize the modified Feynman rules:
\begin{enumerate}
\item{Each line has a polarity which can be either ``$+$'' or ``$-$''.}
\item{Vertices, including counterterms, are either all $+$ or all $-$.}
\item{A $+$ vertex is the familiar one of the in-out formalism, whereas a $-$ 
vertex is its complex conjugate.}
\item{External lines from time-ordered operators are $+$, whereas external 
lines from anti-time-ordered operators are $-$.}
\item{Propagators can be $++$, $+-$, $-+$ or $--$. For our theory these are 
all obtained from the Feynman propagators (\ref{Delta}) and (\ref{fermp}) by 
replacing the conformal coordinate interval $\Delta x^2(x;x')$ with the 
appropriately polarized interval,
\begin{eqnarray}
\Delta x^2_{\scriptscriptstyle ++}(x;x') & \equiv & \Vert \vec{x} \!-\! \vec{x}'
\Vert^2 - (\vert \eta \!-\! \eta'\vert \!-\! i \delta)^2 \; , \\
\Delta x^2_{\scriptscriptstyle +-}(x;x') & \equiv & \Vert \vec{x} \!-\! \vec{x}'
\Vert^2 - (\eta \!-\! \eta' \!+\! i \delta)^2 \; , \\
\Delta x^2_{\scriptscriptstyle -+}(x;x') & \equiv & \Vert \vec{x} \!-\! \vec{x}'
\Vert^2 - (\eta \!-\! \eta' \!-\! i \delta)^2 \; , \\
\Delta x^2_{\scriptscriptstyle --}(x;x') & \equiv & \Vert \vec{x} \!-\! \vec{x}'
\Vert^2 - (\vert \eta \!-\! \eta'\vert \!+\! i \delta)^2 \; .
\end{eqnarray}}
\end{enumerate}

Because the operators in (\ref{test}) are all time-ordered, the associated
external lines have $+$ polarity. At lowest order they can connect to either
a $+$ or a $-$ vertex. Hence the Schwinger-Keldysh result for (\ref{test}) is,
\begin{eqnarray}
\lefteqn{\Bigl\langle \Omega \Bigl\vert T\Bigl[\varphi_{\rm r}(x) \overline{
\psi}_{\rm r}(x) \psi_{\rm r}(x)\Bigr] \Bigr\vert \Omega \Bigr\rangle } 
\nonumber \\
& & = +i f \!\!\int \!\! d^Dx' {a'}^D \left\{ \matrix{i\Delta_{
\scriptscriptstyle ++}(x;x') i \Bigl[{}_i S_j\Bigr]_{\scriptscriptstyle 
++}\!\!\!\!(x;x') i \Bigl[{}_j S_i\Bigr]_{\scriptscriptstyle ++}\!\!\!\!(x';x) 
\cr - i\Delta_{\scriptscriptstyle +-}(x;x') i \Bigl[{}_i S_j\Bigr]_{
\scriptscriptstyle +-}\!\!\!\!(x;x') i \Bigl[{}_j S_i\Bigr]_{\scriptscriptstyle 
-+}\!\!\!\!(x';x)} \right\} + O(f^3) . \quad
\end{eqnarray}
We will ignore the distinction between $++$ and $+-$ until it becomes
significant. In view of (\ref{fermp}) the spinor trace gives,
\begin{equation}
i \Bigl[{}_i S_j\Bigr](x;x') i \Bigl[{}_j S_i\Bigr](x';x) = 
-\frac{\Gamma^2(\frac{D}2)}{4 \pi^D} \frac{\gamma^{\mu}_{ij} 
\gamma^{\nu}_{ji} \Delta x_{\mu} \Delta x_{\nu}}{(a a')^{D-1} \Delta x^{2D}} =
- \frac{\Gamma^2(\frac{D}2)}{\pi^D} \frac{(a a')^{1-D}}{\Delta x^{2D-2}} \; .
\end{equation}
And the lowest order contribution to the VEV takes the form,
\begin{eqnarray}
\lefteqn{\Bigl\langle \Omega \Bigl\vert \varphi_{\rm r}(x) \overline{\psi}_{
\rm r}(x) \psi_{\rm r}(x) \Bigr\vert \Omega \Bigr\rangle } \nonumber \\
& & = -i f \frac{\Gamma^2(\frac{D}2) }{\pi^D a^{D-1}} \!\!\int \!\! d^Dx' a' 
\Biggl\{\frac{i \Delta_{\scriptscriptstyle ++}(x;x')}{\Delta x_{
\scriptscriptstyle ++}^{2D-2}} - \frac{i \Delta_{\scriptscriptstyle +-}(x;x')
}{\Delta x_{\scriptscriptstyle +-}^{2D-2}} \Biggr\} + O(f^3) . \label{trace}
\end{eqnarray}

Here is where the analysis becomes tedious. We have accordingly moved
the technical details to the Appendix, and will here quote only the divergent
parts of the result. The superficially most singular contribution to 
(\ref{trace}) derives from the first line of the scalar propagator 
(\ref{Delta}). We call it $A$,
\begin{eqnarray}
A(x) &\!\!\! \equiv \!\!\!& -i f \frac{\Gamma^2(\frac{D}2) \Gamma(\frac{D}2 
\!-\! 1)}{4\pi^{\frac32 D}} a^{2 -\frac{3D}2} \!\!\int \!\! d^Dx' (a')^{2 -
\frac{D}2} \Biggl\{\frac1{\Delta x_{\scriptscriptstyle ++}^{3D-4}} - 
\frac1{\Delta x_{\scriptscriptstyle +-}^{3D-4}} \Biggr\} , \qquad \label{A} \\
& \!\!\!=\!\!\! & {\rm finite} \label{Ares} \; .
\end{eqnarray}

The next most singular contributions come from the second line of 
(\ref{Delta}),
\begin{eqnarray}
\lefteqn{B_1(x) \equiv -i f H^{D-2} \frac{\Gamma(\frac{D}2) 
\Gamma(D \!-\! 1) }{2^D \pi^{\frac32 D} a^{D-1}} } \nonumber \\
& & \hspace{.2cm} \times \frac{D}{D \!-\! 4} \frac{\Gamma^2(\frac{D}2)}{
\Gamma(D \!-\! 1)} \Bigl(\frac{4}{H^2 a}\Bigr)^{\frac{D}2 - 2} \!\!\int \!\! 
d^Dx' (a')^{3 -\frac{D}2} \Biggl\{\frac1{\Delta x_{\scriptscriptstyle ++}^{
3D-6}} - \frac1{\Delta x_{\scriptscriptstyle +-}^{3D-6}} \Biggr\} , 
\qquad \label{B1} \\
& & =  \frac{f H^4 \mu^{-2\epsilon}}{8 \pi^{4-\epsilon}} \frac{(1 \!-\!
\frac{\epsilon^2}{16}) (1 \!-\! \frac{\epsilon^2}4) \Gamma^2(2 \!-\! 
\frac{\epsilon}2)}{(1 \!-\! \frac34 \epsilon) (1 \!-\! \epsilon) (1 \!-\! 
\frac32 \epsilon)} \Bigl\{-\frac1{2\epsilon^2} \!-\! \frac{\ln(a)}{\epsilon}
\Bigr\} + {\rm finite} , \label{B1res} \\
\lefteqn{B_2(x) \equiv -i f H^{D-2} \frac{\Gamma(\frac{D}2) \Gamma(D \!-\!
1)}{2^D \pi^{\frac32 D} a^{D-1}} } \nonumber \\
& & \hspace{.2cm} \times -\pi \cot\Bigl(\frac{\pi}2 D\Bigr) \!\!\int \!\! d^Dx'
a' \Biggl\{\frac1{\Delta x_{\scriptscriptstyle ++}^{2D-2}} - \frac1{\Delta 
x_{\scriptscriptstyle +-}^{2D-2}} \Biggr\} , \label{B2} \\
& & = \frac{f H^{4 - \epsilon} \mu^{-\epsilon}}{8 \pi^{4- \epsilon} } 
\frac{\pi\epsilon}2 \cot\Bigl(\frac{\pi \epsilon}2\Bigr) \Gamma(1 \!-\! 
\epsilon) \Bigl\{\frac1{\epsilon^2} \!+\! \frac{\ln(a)}{\epsilon} \Bigr\} +
{\rm finite} , \label{B2res} \\
\lefteqn{B_3(x) \equiv -i f H^{D-2} \frac{\Gamma(\frac{D}2) \Gamma(D \!-\!
1) }{2^D \pi^{\frac32 D} a^{D-1}} \!\! \times \!\!\!\!\int \!\! d^Dx' a' 
\ln(a a') \Biggl\{ \!\! \frac1{\Delta x_{\scriptscriptstyle ++}^{2D-2}} \!-\!
\frac1{\Delta x_{\scriptscriptstyle +-}^{2D-2}} \!\! \Biggr\} , \quad \qquad}
\label{B3} \\
& & = \frac{f H^{4 - \epsilon} \mu^{-\epsilon}}{8 \pi^{4- \epsilon} } 
\Gamma(1 \!-\! \epsilon) \Bigl\{\frac3{4 \epsilon} + \frac{\ln(a)}{\epsilon} 
\Bigr\} + {\rm finite} . \label{B3res}
\end{eqnarray}
Only the $n=1$ term from the third line makes a nonzero contribution
in the limit of $D=4$,
\begin{eqnarray}
\lefteqn{C_1(x) \equiv -\frac{i f H^{D-2} \Gamma^2(\frac{D}2)}{2^D 
\pi^{\frac32 D} a^{D-1}} } \nonumber \\
& & \hspace{1.2cm} \times \frac{\Gamma(\frac{D}2 \!+\! 2)}{D \!-\! 6}
\Bigl(\frac{H^2 a}4\Bigr)^{3-\frac{D}2} \!\!\int \!\! d^Dx' 
(a')^{4 -\frac{D}2} \Biggl\{\frac1{\Delta x_{\scriptscriptstyle ++}^{3D-8}} - 
\frac1{\Delta x_{\scriptscriptstyle +-}^{3D-8}} \Biggr\} , \qquad \label{CC1}\\
& & = \frac{f H^4 \mu^{-2\epsilon}}{8 \pi^{4-\epsilon}} \, \frac{(1 \!-\!
\frac{\epsilon}6) (1 \!-\! \frac{\epsilon}4) (1\!-\! \frac{\epsilon}2) 
\Gamma^2(2 \!-\! \frac{\epsilon}2)}{(1 \!-\! \frac32 \epsilon) (1 \!+\! 
\epsilon)} \, \Bigl\{\frac3{8 \epsilon} \Bigr\} + {\rm finite} ,\label{C1res} \\
\lefteqn{C_2(x) \equiv -\frac{i f H^{D-2} \Gamma^2(\frac{D}2)}{2^D 
\pi^{\frac32 D} a^{D-1}} \!\! \times \!\! \frac{\Gamma(D)}{\Gamma(\frac{D}2 
\!+\! 1)} \frac{H^2 a}4 \!\!\int \!\! d^Dx' a^{\prime 2} \Biggl\{ \!\! \frac1{
\Delta x_{\scriptscriptstyle ++}^{2D-4}} \!-\! \frac1{\Delta x_{
\scriptscriptstyle +-}^{2D-4}} \!\! \Biggr\} , \qquad \label{CC2}} \\
& & = \frac{f H^{4-\epsilon} \mu^{-\epsilon}}{2^{3-\epsilon} \pi^{4-\epsilon}}
\, \frac{(1 \!-\! \frac{\epsilon}3) (1\!-\! \frac{\epsilon}2)^2 \Gamma(2 \!-\! 
\epsilon)}{(1 \!-\! \frac{\epsilon}4) (1 \!-\! \epsilon)} \, 
\Bigl\{-\frac3{4 \epsilon}\Bigr\} + {\rm finite} . \label{C2res}
\end{eqnarray}
The $n = 2$ term goes like $\Delta x^4$, which is enough to make denominator 
integrable in $D=4$. We can therefore take $D=4$ for these terms, at which 
point one sees that they vanish. The same argument applies to all terms with 
$n \geq 2$.

The sum of (\ref{Ares}), (\ref{B1res}), (\ref{B2res}), (\ref{B3res}), 
(\ref{C1res}) and (\ref{C2res}) gives,
\begin{equation}
\Bigl({\rm divergent\ constant}\Bigr) + \frac{f H^4}{8 \pi^4} \frac{\ln(a)}{
\epsilon} + {\rm finite} \; , \label{pertpred}
\end{equation}
in perfect agreement with the stochastic prediction (\ref{stochpred})! It is
worth noting that the only the $B$-terms contribute divergent infrared 
logarithms, and those from $B_1$ and $B_2$ cancel. So the result seems to
derive entirely from $B_3$, which itself originated from the explicit factor
of $\ln(a a')$ on the second line of the scalar propagator (\ref{Delta}).

Although it is not really necessary for our purpose of checking the stochastic 
formalism, we remark that the reason the coincident vertex diverges is that 
ordinary renormalization does not generally suffice for composite operators 
such as (\ref{test}). To renormalize local composite operators one must allow 
them to mix with operators of the same or lower dimension. There are 
three local, dimension four operators that can mix with $\varphi_{\rm r}(x)
\overline{\psi}_{\rm r}(x) \psi_{\rm r}(x)$,
\begin{equation}
R \varphi^2_{\rm r} \qquad , \qquad \partial_{\mu} \varphi_{\rm r} \partial_{
\nu} \varphi_{\rm r} g^{\mu\nu} \qquad {\rm and} \qquad \overline{\psi}_{\rm r}
e^{\mu}_{~b} \gamma^b \mathcal{D}_{\mu} \psi_{\rm r} \; .
\end{equation}
The expectation value of the last term vanishes in dimensional regularization. 
The expectation value's of the first two are,
\begin{eqnarray}
\Bigl\langle \Omega \Bigl\vert T\Bigl[R \varphi^2_{\rm r}(x) \Bigr]\Bigr\vert 
\Omega \Bigr\rangle = \frac{H^{4-\epsilon}}{2^{3-\epsilon} \pi^{2-\frac{
\epsilon}2}} \frac{\Gamma(5 \!-\! \epsilon)}{\Gamma(2 \!-\! \frac{\epsilon}2)} 
\Bigl\{ \frac{\pi \epsilon}2 \cot\Bigl(\frac{\pi \epsilon}2 \Bigr) + \ln(a)
\Bigr\} , \qquad \\
\Bigl\langle \Omega \Bigl\vert T^*\Bigl[\partial_{\mu} \varphi_{\rm r}(x)
\partial_{\nu} \varphi_{\rm r}(x) g^{\mu\nu} \Bigr]\Bigr\vert \Omega 
\Bigr\rangle = \frac{H^{4-\epsilon}}{2^{5-\epsilon} \pi^{2-\frac{\epsilon}2}} 
\frac{(4 \!-\! \epsilon) \Gamma(3 \!-\! \epsilon)}{\Gamma(2 \!-\! 
\frac{\epsilon}2)} .
\end{eqnarray}
We can choose the coefficient of $R \varphi^2_{\rm r}(x)$ to completely cancel 
the second term of (\ref{pertpred}). Then we can choose the coefficient
of $\partial_{\mu} \varphi_{\rm r}(x) \partial_{\nu} \varphi_{\rm r}(x) 
g^{\mu\nu}$ to cancel whatever constant terms remain.

\section{Stochastic Stress Tensor}

To understand how this model sources gravitational back-reaction we must
study the Yukawa stress tensor,
\begin{eqnarray}
T_{\mu\nu} & \equiv & -\frac1{\sqrt{-g}} e_{(\mu b} \frac{\delta S}{
\delta e^{\nu) b}} \; , \\
& = & -\frac{i}2 \Bigl[ \overline{\psi} e_{(\mu b} \gamma^b \mathcal{D}_{\nu)} 
\psi \!-\! \overline{\psi} \, \overleftarrow{\mathcal{D}}_{(\mu} e_{\nu) b} 
\gamma^b \psi\Bigr] + \partial_{\mu} \varphi \partial_{\nu} \varphi - \frac12 
g_{\mu\nu} g^{\rho\sigma} \partial_{\rho} \varphi \partial_{\sigma} \varphi 
\nonumber \\
& & \hspace{1cm} + \delta \xi \Big[(R_{\mu\nu} \!-\! \frac12 g_{\mu\nu} R) 
\varphi^2 \!+\! g_{\mu\nu} (\varphi^2)^{;\rho}_{~~\rho} \!-\! (\varphi^2)_{
;\mu\nu}\Bigr] - \frac{\delta \lambda}{4!} \varphi^4 g_{\mu\nu} \; . \qquad 
\end{eqnarray}
Integrating out the fermions converts the fermionic terms to a purely scalar 
expression that we can evaluate for constant field configurations using 
(\ref{C2}) and (\ref{C3}),
\begin{eqnarray}
\lefteqn{e^{-i \Gamma[\varphi]} \Fint [d\overline{\psi}] [d\psi] e^{i 
S[\varphi,\overline{\psi},\psi]} \times \frac{i}{2 a} \Bigl[\overline{\psi}
\gamma_{(\mu} \mathcal{D}_{\nu)} \psi - \overline{\psi} \, \overleftarrow{
\mathcal{D}}_{(\mu} \gamma_{\nu)} \psi \Bigr] } \nonumber \\
& & \hspace{0cm} = -\frac{i a}2 \lim_{x' \rightarrow x} {\rm Tr}\Biggl[ -
\gamma_{(\mu} \mathcal{D}_{\nu} i \mathcal{S}[f\varphi](x;x') + i \mathcal{S}[
f\varphi](x;x') \overleftarrow{ \mathcal{D}}_{(\mu}' \gamma_{\nu)} \Biggr] 
\; , \qquad \\
& & \hspace{0cm} \longrightarrow \frac{4 H^{D}}{(4 \pi)^{\frac{D}2}}
\frac{\Gamma(1 \!-\! \frac{D}2)}{D} \Biggl\Vert \frac{\Gamma(\frac{D}2 \!+\! i
\frac{f \varphi}{H})}{\Gamma(1 \!+\! i \frac{f\varphi}{H})} \Biggr\Vert^2 
\Bigl(\frac{f \varphi}{H}\Bigr)^2 g_{\mu\nu} \; , \qquad \label{fterm} \\
& & \hspace{0cm} = \frac{H^D}{(4\pi)^{\frac{D}2}} \Biggl\{ \frac4{D} 
\Gamma\Bigl(1 \!-\! \frac{D}2\Bigr) \Bigl[\Bigl(\frac{f \varphi}{H}\Bigr)^2
\!+\! \Bigl(\frac{f \varphi}{H}\Bigr)^4\Bigr] + 2 \Bigl(\frac{f \varphi}{H}
\Bigr)^2 \nonumber \\
& & \hspace{1.5cm}+ \Bigl[\Bigl(\frac{f \varphi}{H}\Bigr)^2 \!+\! \Bigl(\frac{f 
\varphi}{H}\Bigr)^4\Bigr] \Bigl[\psi\Bigl(1 \!+\! i\frac{f \varphi}{H}\Bigr)
+ \psi\Bigl(1 \!-\! i\frac{f \varphi}{H}\Bigr)\Bigr] + O(\epsilon) \Biggr\} 
g_{\mu\nu} . \qquad
\end{eqnarray}
Because the differentiated fields in $T_{\mu\nu}$ cannot contribute leading 
order logarithms we see that the stress tensor takes the form,
\begin{equation}
T_{\mu\nu} \longrightarrow -V_{\rm s}(\varphi) g_{\mu\nu} \; ,
\end{equation}
where the potential is,
\begin{eqnarray}
\lefteqn{V_{\rm s}(\varphi) = \frac{H^4}{8 \pi^2} \Biggl\{ \Bigl[\frac12 \!-\! 
\gamma\Bigr] \Bigl(\frac{f \varphi}{H}\Bigr)^2 + \Bigl[\frac14 \!-\! \gamma 
\!+\! \zeta(3) \Bigr] \Bigl(\frac{f \varphi}{H}\Bigr)^4 } \nonumber \\
& & \hspace{3cm} - \frac12 \Bigl[\Bigl(\frac{f \varphi}{H}\Bigr)^2 \!+\! 
\Bigl(\frac{f \varphi}{H}\Bigr)^4\Bigr] \Bigl[\psi\Bigl(1 \!+\! i\frac{f 
\varphi}{H}\Bigr) + \psi\Bigl(1 \!-\! i\frac{f \varphi}{H}\Bigr)\Bigr] 
\Biggr\} , \qquad \label{gravpot} \\
& & \hspace{-.7cm} = \frac{H^4}{8 \pi^2} \Biggl\{\frac12 \Bigl(\frac{f 
\varphi}{H}\Bigr)^2 \!+\! \frac14 \Bigl(\frac{f \varphi}{H}\Bigr)^4 
\!-\! \sum_{n=2}^{\infty} (-1)^n \Bigl[\zeta(2n \!-\! 1) \!-\!  \zeta(2n 
\!+\!  1)\Bigr] \Bigl(\frac{f \varphi}{H} \Bigr)^{2n + 2} \Biggr\} . \qquad
\end{eqnarray}

This is not the same potential $V(\varphi)$ we found in section 3! Unlike
that potential, $V_{\rm s}(\varphi)$ has positive curvature at $\varphi \!=\!
0$. However, the leading asymptotic behavior for large $\varphi$ is the same,
\begin{eqnarray}
\lefteqn{V_{\rm s}(\varphi) = -\frac{H^4}{8 \pi^2} \Biggl\{\Bigl(\frac{f 
\varphi}{H}\Bigr)^4 \ln\Bigl(\frac{f \vert \varphi\vert}{H}\Bigr) - \Bigl[
\zeta(3) \!+\! \frac14 \!-\! \gamma\Bigr] \Bigl(\frac{f \varphi}{H}\Bigr)^4 } 
\nonumber \\
& & \hspace{1.5cm} + \Bigl(\frac{f \varphi}{H}\Bigr)^2 \ln\Bigl(\frac{f \vert
\varphi\vert}{H}\Bigr) - \Bigl(\frac5{12} \!-\! \gamma\Bigr) \Bigl(\frac{f 
\varphi}{H}\Bigr)^2 + O\Bigl(\ln\Bigl[\frac{f \vert \varphi\vert}{H}\Bigr]
\Bigr) \Biggr\} .
\end{eqnarray}
The evolution of $\varphi$ is controlled by $V(\varphi)$. So the scalar rolls
away from $\varphi \!=\! 0$, even though this initially means {\it moving up} 
the potential $V_{\rm s}(\varphi)$. 

What $V_{\rm s}(\varphi)$ gives is the 
expectation value of the operator which is the source of gravitational
back-reaction. The two potentials disagree because the scalar effective 
potential $V(\varphi)$ involves the square of the Hubble parameter, which is 
really $H^2 \longrightarrow R/12$ for a general metric.\footnote{This nonlinear
dependence upon the Ricci scalar does not imply the usual kinetic instability
associated with higher derivative gravity \cite{RPW4}.} That has consequences
for the way $V(\varphi)$ sources gravity. One can see this from the familiar
case of a conformal coupling term in the Lagrangian, $-\frac12 \varphi^2 R 
\sqrt{-g}$. In 4-dimensional de Sitter the Ricci scalar is $R \!=\! 12 H^2$, 
so the contribution to the scalar effective potential is,
\begin{equation}
\Delta \mathcal{L} = -\frac12 \varphi^2 R \sqrt{-g} \qquad \Longrightarrow 
\qquad \Delta V(\varphi) = 6 H^2 \varphi^2 \; .
\end{equation}
But the stress tensor takes account of the way the Ricci scalar depends
upon the metric for a general geometry,
\begin{equation}
\Delta \mathcal{L} = -\frac12 \varphi^2 R \sqrt{-g} \quad \Longrightarrow 
\quad \Delta T_{\mu\nu} = \Bigl(R_{\mu\nu} \!-\! \frac12 g_{\mu\nu} R\Bigr) 
\varphi^2 + g_{\mu\nu} (\varphi^2)^{;\rho}_{~~\rho} - (\varphi^2)_{;\mu\nu}\; .
\end{equation}
Differentiated scalars cannot contribute leading order infrared logarithms, 
and the de Sitter Einstein tensor is $R_{\mu\nu} \!-\! \frac12 g_{\mu\nu} R
\!=\! -3 H^2 g_{\mu\nu}$, so the induced potential in the stress tensor is,
\begin{equation}
\Delta \mathcal{L} = -\frac12 \varphi^2 R \sqrt{-g} \qquad \Longrightarrow 
\qquad \Delta V_{\rm s}(\varphi) = 3 H^2 \varphi^2 \; .
\end{equation}

\section{Discussion}

Any theory which includes either massless minimally coupled scalars or 
gravitons will show infrared logarithms in the expectations values of certain
operators. These enhance loop effects by powers of $\ln[a(t)]$, where $a(t) 
\sim e^{H t}$ is the inflationary scale factor. Of course loop effects are 
still down by powers of the (presumed small) loop counting parameter, but 
continued evolution must eventually bring about a situation in which the
factors of $\ln[a(t)]$ overcome the small loop counting parameter and
cause the breakdown of perturbation theory. Starobinski\u{\i} has long
advocated gaining control over this nonperturbative regime by studying the
series comprised of the leading infrared logarithm at each loop order
\cite{AAS}. In fact he has completely solved for the leading log limit of
a massless, minimally coupled scalar with arbitrary potential \cite{SY}.

In scalar potential models every field is ``active.'' That is, it can
potentially contribute to an infrared logarithm. Because the leading log
result requires that {\it all} fields at a given order contribute infrared
logarithms, one can perform an infrared truncation on the fields. This is 
why Starobinski\u{\i}'s formalism ends up being so wonderfully simple.
More general theories also possess ``passive'' fields which cannot
themselves contribute to an infrared logarithm. However, these passive
fields can propagate infrared logarithms obtained from interaction with 
active fields. They can also mediate interactions between active fields.
Differentiated active fields play much the same role. 

In propagating infrared logarithms, and mediating interactions between
active fields, the ultraviolet parts of the passive fields (and differentiated 
actives) contribute on an equal footing with the infrared. It is therefore 
invalid to infrared truncate either passive fields or differentiated active
fields which appear in interactions. The correct procedure instead is to
formally integrate out the passive fields, and the differentiated active
fields, both from the action and from whatever operator is being studied. The
expression which results is generally not local, but it contains only active 
fields. Because the nonlocality is confined to inverse differential operators
of passive fields, which {\it cannot} cause infrared logarithms, the 
associated Green's functions are always dominated by positive powers of the 
scale factor whose explosive growth weights the result completely at the upper 
limit in the leading logarithm approximation. Hence the nonlocal effective 
action degenerates to a completely local and computable effective potential.
At this point one has a local potential model of the form Starobinski\u{\i}
has already solved {\it in toto} \cite{AAS,SY}.

Massless Yukawa theory is a wonderfully simple testing ground for these
ideas because it contains a passive field --- the fermion --- without any 
differentiated active fields.\footnote{Yukawa theory is so much simpler 
than SQED that this paper was complete well before a very similar analysis
of SQED which was begun at approximately the same time 
\cite{PTsW1,PTsW2,PTsW3}.} It also lacks the subtle gauge fixing problems 
of SQED \cite{KW} and gravity \cite{TW5,MW1}. In this paper we have 
exploited the classic solution of Candelas and Raine \cite{CR} to derive 
the Yukawa stochastic effective potential $V(\varphi)$ (\ref{psiform}). We 
have checked the technique with an explicit two loop computation of the 
coincident vertex function. The result is in perfect agreement with the 
stochastic prediction.

We have also obtained a leading log result for the stress tensor as a 
function of the scalar. Although this stress tensor takes the form $-
g_{\mu\nu} V_{\rm s}(\varphi)$, our result for $V_{\rm s}(\varphi)$ 
(\ref{gravpot}) it is not quite the same as the effective potential 
$V(\varphi)$ (\ref{psiform}) which governs the scalar's evolution. The 
reason for this is that both potentials depend upon the dimensionless 
quantity, $(f \varphi/H)^2$, and the factors of $H^2 = R/12$ in this exert a 
nontrivial influence upon the way in which this model sources gravity.
So one determines the scalar's evolution using $V(\varphi)$, and one
finds its impact upon gravity from $V_{\rm s}(\varphi)$. The two potentials
differ, but they are each correct.

A curious and potentially significant feature of both potentials is
that they are unbounded below. The physics behind this seems to be 
very solid: inflationary particle production drives the scalar away
from zero, which induces a fermion mass. That increases the magnitude 
of the fermion 0-point energy, which makes for a negative effective 
potential because fermion vacuum energy is negative. We note that 
scalars seem always to induce growing mass \cite{PW2,PW3,PP,GP}, so we expect 
that the effective potential of SQED will be positive for large fields. 
By comparison, gravitons seem to induce a growing field strength 
renormalization \cite{MW2}. It is intriguing to speculate on what that 
might mean for back-reaction in theories of gravity plus matter.

Because the scalar effective potential is unbounded below, this model
should decay forever. However, Hubble friction will make the evolution 
dominated by inflationary particle production until the scalar reaches
nonperturbatively large values. Although the initial effect is to raise
the gravitating energy density, the large field results for $V(\varphi)$
and $V_{\rm s}(\varphi)$ agree. It seems inevitable that pure Yukawa
must be unstable against slow decay to anti-de Sitter. In the Standard
Model of flat space this same tendency is controlled by the positive
effective potential from the gauge bosons. One naturally wonders what
the result might be for inflation, and whether or not this might be
parleyed into a model in which inflation might be gotten to end {\it 
without} endless decay into anti-de Sitter.

\section{Appendix: Integrals from Section 4}

In section 4 we reduced the perturbative expression for the two loop
contribution to (\ref{test}) to a sum of dimensionally regulated integrals
(\ref{A}), (\ref{B1}), (\ref{B2}), (\ref{B3}), (\ref{CC1}) and (\ref{CC2}).
The next step is to partially integrate the inverse powers of $\Delta x^2$
until they become integrable in $D=4$. There is no distinction between $++$
and $+-$ terms at this stage. The identities we need are,
\begin{eqnarray}
\frac1{\Delta x^{3D-8}} \!\!\!\!\! & = & \!\!\!\!\! \frac{\partial^2}{(3D \!-\!
10) 2 (D \!-\!4)} \Bigl( \frac1{\Delta x^{3D-10}} \Bigr) , \\
\frac1{\Delta x^{3D-6}} \!\!\!\!\! & = & \!\!\!\!\! \frac{\partial^4}{(3D \!-\!
8) (3D \!-\! 10) 4 (D \!-\!3) (D \!-\!4)} \Bigl( \frac1{\Delta x^{3D-10}} 
\Bigr) , \\
\frac1{\Delta x^{3D-4}} \!\!\!\!\! & = & \!\!\!\!\! \frac{\partial^6}{(3D \!-\!
6) (3D \!-\! 8) (3D \!-\! 10) 8 (D \!-\! 2) (D \!-\!3) (D \!-\!4)} \Bigl( 
\frac1{\Delta x^{3D-10}} \Bigr) , \qquad \\
\frac1{\Delta x^{2D-4}} \!\!\!\!\! & = & \!\!\!\!\! \frac{\partial^2}{(2D \!-\!
6) (D \!-\!4)} \Bigl( \frac1{\Delta x^{2D-6}} \Bigr) , \\
\frac1{\Delta x^{2D-2}} \!\!\!\!\! & = & \!\!\!\!\! \frac{\partial^4}{(2D \!-\!
4) (2D \!-\! 6) (D \!-\!2) (D \!-\!4)} \Bigl( \frac1{\Delta x^{2D-6}} \Bigr) .
\end{eqnarray}

Because we are integrating over $x^{\prime \mu}$, derivatives with respect 
to $x^{\mu}$ can be taken outside the integral, leaving an integrand which is
finite in $D=4$ dimensions. The limit $D \!=\! 4$ could be taken at this point 
except for the factors of $1/(D-4)$ which were picked up from the last partial 
integration. To segregate the divergence on a local term we add zero in the 
form,
\begin{equation}
\partial^2 \Bigl(\frac1{\Delta x_{\scriptscriptstyle ++}^{D-2}}\Bigr) -
\frac{i 4 \pi^{\frac{D}2}}{\Gamma(\frac{D}2 \!-\! 1)} \, \delta^D(x - x') =
0 = \partial^2 \Bigl(\frac1{\Delta x_{\scriptscriptstyle +-}^{D-2}}\Bigr) \; .
\end{equation}
Once this has been added we can take the limit $D=4$ in the nonlocal term.

We will work this out for the $+-$ term,
\begin{eqnarray}
\lefteqn{\frac{\partial^2}{2D \!-\! 8} \Biggl\{ \frac1{\Delta x_{
\scriptscriptstyle +-}^{3D-10}} \Biggr\} = \frac{\partial^2}{2D \!-\! 8} 
\Biggl\{ \frac1{\Delta x_{\scriptscriptstyle +-}^{3D-10}} - 
\frac{\mu^{2D-8}}{\Delta x_{\scriptscriptstyle +-}^{D-2}}\Biggr\} , } \\
& & = \frac{\partial^2}{2D \!-\! 8} \Biggl\{ \frac{\mu^{3(D-4)}}{\Delta
x^2_{\scriptscriptstyle +-}} \Biggl[ \Bigl(\mu^2 \Delta x^2_{
\scriptscriptstyle +-}\Bigr)^{-\frac32 (D-4)} - \Bigl(\mu^2 \Delta x^2_{
\scriptscriptstyle +-}\Bigr)^{-\frac12 (D-4)} \Biggr] \Biggr\} , \\
& & = \mu^{3(D-4)} \partial^2 \Biggl\{ \frac{-\frac12 \ln\Bigl( \mu^2 \Delta 
x_{\scriptscriptstyle +-}^2\Bigr) \!+\! \frac{(D \!-\! 4)}2 \ln^2\Bigl( \mu^2 
\Delta x_{\scriptscriptstyle +-}^2\Bigr) \!+\! O\Bigl((D\!-\! 4)^2 \Bigr)}{
\Delta x_{\scriptscriptstyle +-}^2} \Biggr\} . \qquad \label{3DR}
\end{eqnarray}
The other result $+-$ we need is,
\begin{eqnarray}
\lefteqn{\frac{\partial^2}{D \!-\! 4} \Biggl\{ \frac1{\Delta x_{
\scriptscriptstyle +-}^{2D-6}} \Biggr\} = \frac{\partial^2}{D \!-\! 4} 
\Biggl\{ \frac1{\Delta x_{\scriptscriptstyle +-}^{2D-6}} -
\frac{\mu^{D-4}}{\Delta x_{\scriptscriptstyle +-}^{D-2}}\Biggr\} , } \\
& & \hspace{-.5cm} = \mu^{2(D-4)} \partial^2 \Biggl\{ \frac{-\frac12 \ln\Bigl( 
\mu^2 \Delta x_{\scriptscriptstyle +-}^2\Bigr) \!+\! \frac{3(D \!-\! 4)}{8} 
\ln^2\Bigl(\mu^2 \Delta x_{\scriptscriptstyle +-}^2\Bigr) \!+\! O\Bigl((D
\!-\! 4)^2 \Bigr)}{\Delta x_{\scriptscriptstyle +-}^2} \Biggr\} . \qquad
\label{2DR}
\end{eqnarray}
It is only on account of the explicit factors of $1/(D \!-\! 4)$ in $B_1(x)$ 
and $\cot(\frac{\pi}2 D)$ in $B_2(x)$ that we must keep the order $(D \!-\! 4)$ 
terms in relations (\ref{3DR}) and (\ref{2DR}). The analogous $++$ relations
are,
\begin{eqnarray}
\lefteqn{\frac{\partial^2}{2D \!-\! 8} \Biggl\{ \frac1{\Delta x_{
\scriptscriptstyle ++}^{3D-10}} \Biggr\} = \frac{\mu^{2D-8}}{2D \!-\! 8}
\frac{i 4 \pi^{\frac{D}2}}{\Gamma(\frac{D}2 \!-\! 1)} \, 
\delta^D(x - x') } \nonumber \\
& & + \mu^{3(D-4)} \partial^2 \Biggl\{ \frac{-\frac12 \ln\Bigl( \mu^2 \Delta 
x_{\scriptscriptstyle ++}^2\Bigr) \!+\! \frac{(D \!-\! 4)}2 \ln^2\Bigl( \mu^2 
\Delta x_{\scriptscriptstyle ++}^2\Bigr) \!+\! O\Bigl((D\!-\! 4)^2 \Bigr)}{
\Delta x_{\scriptscriptstyle ++}^2} \Biggr\} . \qquad
\end{eqnarray}
\begin{eqnarray}
\lefteqn{\frac{\partial^2}{D \!-\! 4} \Biggl\{ \frac1{\Delta x_{
\scriptscriptstyle ++}^{2D-6}} \Biggr\} = \frac{\mu^{D-4}}{D \!-\! 4}
\frac{i 4 \pi^{\frac{D}2}}{\Gamma(\frac{D}2 \!-\! 1)} \, 
\delta^D(x - x') } \nonumber \\
& & \hspace{-.5cm} + \mu^{2(D-4)} \partial^2 \Biggl\{ \frac{-\frac12 \ln\Bigl( 
\mu^2 \Delta x_{\scriptscriptstyle ++}^2\Bigr) \!+\! \frac{3(D \!-\! 4)}{8} 
\ln^2\Bigl(\mu^2 \Delta x_{\scriptscriptstyle ++}^2\Bigr) \!+\! O\Bigl((D
\!-\! 4)^2 \Bigr)}{\Delta x_{\scriptscriptstyle ++}^2} \Biggr\} . \qquad
\end{eqnarray}

The use of these partial integration identities results in each of the
terms (\ref{A}), (\ref{B1}), (\ref{B2}), (\ref{B3}), (\ref{CC1}) and
(\ref{CC2}) having a finite, nonlocal part and a potentially divergent,
local part. For $A(x)$ these are,
\begin{eqnarray}
A_N(x) & = & \frac{if a^{-4}}{2^{10} 3 \pi^6} \partial^6 \int d^4x'
\Biggl\{\frac{\ln\Bigl(\mu^2 \Delta x^2_{\scriptscriptstyle ++}\Bigr)}{
\Delta x^2_{\scriptscriptstyle ++}} - \frac{\ln\Bigl(\mu^2 \Delta x^2_{
\scriptscriptstyle +-}\Bigr)}{\Delta x^2_{\scriptscriptstyle +-}}\Biggr\} ,\\
A_L(x) & = & \frac{f \Gamma^2(\frac{D}2)}{\pi^D}
\frac{\mu^{2D-8} a^{2-\frac{3D}2} \partial^4 a^{2-\frac{D}2}}{(3D \!-\!
6) (3D \!-\! 8) (3D \!-\! 10) (2D \!-\! 4) (2D \!-\!6) (2D \!-\!8)} . \qquad
\end{eqnarray}
Before giving the nonlocal and local terms for the $B_i(x)$ and $C_i(x)$
we will evaluate the two above. The local term is simple because the derivatives
in the numerator are so easy,
\begin{equation}
a^{2-\frac{3D}2} \partial^4 a^{2-\frac{D}2} = \Bigl(2-\frac{D}2\Bigr)
\Bigl(3-\frac{D}2\Bigr) \Bigl(4-\frac{D}2\Bigr) \Bigl(5-\frac{D}2\Bigr)
H^4 a^{8-2D} \; .
\end{equation}
The factor of $\Bigl(2-\frac{D}2\Bigr)$ means that $A_L(x)$ is actually
finite and we can set $D=4$,
\begin{equation}
A_L(x) \longrightarrow - \frac{f H^4}{2^8 \pi^4} \; .
\end{equation}

To evaluate any of the nonlocal terms it is best to extract two more 
derivatives,
\begin{equation}
A_N(x) = \frac{if a^{-4}}{2^{13} 3 \pi^6} \partial^8 \int d^4x'
\left\{ \matrix{\ln^2\Bigl(\mu^2 \Delta x^2_{\scriptscriptstyle ++}\Bigr)
- 2 \ln\Bigl(\mu^2 \Delta x^2_{\scriptscriptstyle ++}\Bigr) \cr
- \ln^2\Bigl(\mu^2 \Delta x^2_{\scriptscriptstyle +-}\Bigr) + 2 \ln\Bigl(
\mu^2 \Delta x^2_{\scriptscriptstyle +-}\Bigr) } \right\} .
\end{equation}
The point of doing this is that we can now exploit the exact cancellation
between $++$ and $+-$ terms outside the past light-cone. (Note that we do
not want to do this before the last derivative is extracted because the
limits of integration must be constant for us to extract derivatives.) 
We define the temporal and spatial intervals in the obvious ways,
\begin{equation}
\Delta \eta \equiv \eta - \eta' \qquad , \qquad \Delta r \equiv \Vert
\vec{x} - \vec{x}' \Vert \; .
\end{equation}
It is then straightforward to show,
\begin{eqnarray}
\ln\Bigl(\mu^2 \Delta x^2_{\scriptscriptstyle ++}\Bigr) -
\ln\Bigl(\mu^2 \Delta x^2_{\scriptscriptstyle +-}\Bigr) & = & 2 \pi i
\theta( \Delta \eta - \Delta r) \; , \\
\ln^2\Bigl(\mu^2 \Delta x^2_{\scriptscriptstyle ++}\Bigr) -
\ln^2\Bigl(\mu^2 \Delta x^2_{\scriptscriptstyle +-}\Bigr) & = & 4 \pi i
\theta( \Delta \eta - \Delta r) \ln\Bigl[\mu^2 (\Delta \eta^2
- \Delta r^2)\Bigr] . \qquad
\end{eqnarray}
These relations bring the nonlocal term to the form,
\begin{equation}
A_N(x) = \frac{-f a^{-4}}{2^{11} 3 \pi^5} \partial^8 \int_{\eta_I}^{\eta}
\!\!\! d\eta' \int d^3x' \theta(\Delta \eta - \Delta r) \Biggl\{\ln\Bigl[\mu^2 
(\Delta \eta^2 - \Delta r^2)\Bigr] - 1\Biggr\} ,
\end{equation}
where the initial time is $\eta_I \equiv -1/H$. Note that $A_N(x)$ is now 
manifestly real.

The next step is to make the change of variables $\vec{r} = \vec{x}' - \vec{x}$
and perform the angular integrals,
\begin{equation}
A_N(x) = \frac{-f a^{-4}}{2^9 3 \pi^4} \partial_0^8 \int_{\eta_I}^{\eta}
\!\!\! d\eta' \int_0^{\Delta \eta} dr r^2 \Biggl\{\ln\Bigl[\mu^2 (\Delta 
\eta^2 - r^2)\Bigr] - 1\Biggr\} .
\end{equation}
We then make the change of variables $r = \Delta \eta z$ and perform the 
integration over $z$,
\begin{equation}
A_N(x) = \frac{-f a^{-4}}{2^9 3 \pi^4} \partial_0^8 \int_{\eta_I}^{\eta}
\!\!\! d\eta' \Delta \eta^3 \Biggl\{ \frac23 \ln\Bigl(2 \mu \Delta \eta\Bigr)
- \frac{11}9\Biggr\} .
\end{equation}
Owing to the factor of $\Delta \eta^3$, three of the external derivatives
can be brought inside the integral,
\begin{equation}
A_N(x) = \frac{-f a^{-4}}{2^9 3 \pi^4} \partial_0^5 \int_{\eta_I}^{\eta}
\!\!\! d\eta' 4 \ln\Bigl(2 \mu \Delta \eta\Bigr) .
\end{equation}
At this stage one performs the integral over $\eta'$ and acts the derivatives,
\begin{equation}
A_N(x) = -\frac{f a^{-4}}{2^7 3 \pi^4} \partial_0^5 \Biggl\{ {\Delta \eta}_I 
\Bigl[\ln(2 \mu {\Delta \eta}_I) - 1\Bigr] \Biggr\} 
= \frac{f H^4}{2^6 \pi^4} \frac1{(a \!-\! 1)^4} \; .
\end{equation}

For the $B$-terms it is best to convert from $D$ to $\epsilon = 4-D$.
All three of the $B$ terms (\ref{B1}), (\ref{B2}) and (\ref{B3}), contain an 
overall factor of,
\begin{equation}
-i f H^{D-2} \frac{\Gamma(\frac{D}2) \Gamma(D \!-\! 1)}{2^D 
\pi^{\frac32 D} a^{D-1}} = - i f H^{2 - \epsilon} \frac{\Gamma(2 \!-\!
\frac{\epsilon}2) \Gamma(3 \!-\! \epsilon)}{2^{4 - \epsilon} \pi^{6
- \frac32 \epsilon} a^{3-\epsilon}} \; . \label{factor}
\end{equation}
The integrand of the $B_1$ term (\ref{B1}) is this overall factor times,
\begin{eqnarray}
\lefteqn{ \frac{D}{D \!-\! 4} \frac{\Gamma^2(\frac{D}2)}{\Gamma(D \!-\!1)}
\Bigl(\frac14 H^2 a a'\Bigr)^{2 - \frac{D}2} a' \left\{ \frac1{{\Delta x}_{
\scriptscriptstyle ++}^{3D-6}} - \frac1{{\Delta x}_{ \scriptscriptstyle +-}^{
3D-6}} \right\} = \Bigl(\frac{4 \!-\! \epsilon}{-\epsilon}\Bigr) 
\frac{\Gamma^2(2 \!-\! \frac{\epsilon}2)}{\Gamma(3 \!-\! \epsilon)} } 
\nonumber \\
& & \hspace{-.7cm} \times \frac{\Bigl( \frac14 H^2 a a'\Bigr)^{\frac12 
\epsilon} a' \, \partial^2}{(4 \!-\! 3 \epsilon) (2 \!-\! 3 \epsilon) (2 \!-\! 
2 \epsilon) (-2\epsilon)} \!\!\left\{\! \mu^{-3\epsilon} \partial^2 \!\!\left[ 
\frac{\epsilon \ln(\mu^2 {\Delta x}^2_{\scriptscriptstyle ++}) + \epsilon^2 
\ln^2(\mu^2 {\Delta x}^2_{\scriptscriptstyle ++}) + \dots}{{\Delta x}^2_{
\scriptscriptstyle ++}} \right] \right. \nonumber \\
& & \hspace{4cm} \left. - \Bigl( {\scriptscriptstyle ++} \rightarrow 
{\scriptscriptstyle +-}\Bigr) + \frac{i 4 \pi^{2 - \frac12 \epsilon} \mu^{-2 
\epsilon}}{\Gamma(1 \!-\! \frac{\epsilon}2 )} \delta^D(x - x') \right\} .
\end{eqnarray}
The associated local term is,
\begin{eqnarray}
B_{1L}(x) & = & \frac{f H^2 \mu^{-2\epsilon}}{32 \pi^{4-\epsilon}} 
\frac{(1 \!-\!\frac{\epsilon}4) (1 \!-\! \frac{\epsilon}2) \Gamma^2(2 \!-\!
\frac{\epsilon}2)}{(1 \!-\! \frac34 \epsilon) (1 \!-\! \epsilon) (1 \!-\!
\frac32 \epsilon) \epsilon^2} \, a^{-3+\frac32 \epsilon} \partial^2 a^{1 +
\frac12 \epsilon} \; , \\
& = & - \frac{f H^4 \mu^{-2\epsilon}}{16 \pi^{4-\epsilon}} 
\frac{(1 \!-\!\frac{\epsilon^2}{16}) (1 \!-\! \frac{\epsilon^2}4) \Gamma^2(2 
\!-\! \frac{\epsilon}2)}{(1 \!-\! \frac34 \epsilon) (1 \!-\! \epsilon) 
(1 \!-\! \frac32 \epsilon) \epsilon^2} \, a^{2\epsilon} \; .
\end{eqnarray}

The integrand of the $B_2$ term (\ref{B2}) is the factor (\ref{factor})
times,
\begin{eqnarray}
\lefteqn{ -\pi \cot\Bigl(\frac{\pi}2 D\Bigr) a' \left\{ \frac1{{\Delta x}_{
\scriptscriptstyle ++}^{2D-2}} - \frac1{{\Delta x}_{ \scriptscriptstyle +-}^{
2D-2}} \right\} = \pi \cot\Bigl(\frac{\pi \epsilon}2 \Bigr) } \nonumber \\
& & \hspace{-.7cm} \times \frac{a' \, \partial^2}{(4 \!-\! 2 \epsilon) (2 
\!-\! 2 \epsilon) (2 \!-\! \epsilon) (-\epsilon)} \left\{ \mu^{-2\epsilon} 
\partial^2 \!\!\left[ \frac{\frac{\epsilon}2 \ln(\mu^2 {\Delta x}^2_{
\scriptscriptstyle ++}) \!+\! \frac38 \epsilon^2 \ln^2(\mu^2 {\Delta x}^2_{
\scriptscriptstyle ++}) \!+\! \dots}{{\Delta x}^2_{ \scriptscriptstyle ++}} 
\right] \right. \nonumber \\
& & \hspace{4cm} \left. - \Bigl( {\scriptscriptstyle ++} \rightarrow 
{\scriptscriptstyle +-}\Bigr) + \frac{i 4 \pi^{2 - \frac12 \epsilon} \mu^{-
\epsilon}}{\Gamma(1 \!-\! \frac{\epsilon}2 )} \delta^D(x - x') \right\} .
\end{eqnarray}
The appropriate local term for $B_2$ is,
\begin{eqnarray}
B_{2L}(x) & = & \frac{-f H^{2 - \epsilon} \mu^{-\epsilon}}{2^{6 - 
\epsilon} \pi^{4-\epsilon}} \frac{ \pi \cot(\frac{\pi \epsilon}2) \Gamma(3 
\!-\! \epsilon)}{(1 \!-\! \frac{\epsilon}2) (1 \!-\! \epsilon) \epsilon} \, 
a^{-3+\epsilon} \partial^2 a \; , \\
& = & \frac{f H^{4 - \epsilon} \mu^{-\epsilon}}{8 \pi^{4- \epsilon} } 
\frac{\pi\epsilon}2 \cot\Bigl(\frac{\pi \epsilon}2\Bigr) \Gamma(1 \!-\! 
\epsilon) \frac{a^{\epsilon}}{\epsilon^2} .
\end{eqnarray}
The integrand of the $B_3$ term (\ref{B3}) is the factor (\ref{factor})
times,
\begin{eqnarray}
\lefteqn{ a' \ln(a a') \left\{ \frac1{{\Delta x}_{\scriptscriptstyle ++}^{
2D-2}} - \frac1{{\Delta x}_{ \scriptscriptstyle +-}^{2D-2}} \right\} = 
\frac{a' \ln(a a') \, \partial^2}{(4 \!-\! 2 \epsilon) (2 \!-\! 2 \epsilon) 
(2 \!-\! \epsilon) (-\epsilon)} } \nonumber \\
& & \hspace{-.7cm} \times \left\{ \mu^{-2\epsilon} \partial^2 \!\!\left[ 
\frac{\frac{\epsilon}2 \ln(\mu^2 {\Delta x}^2_{\scriptscriptstyle ++}) \!+\! 
\dots}{{\Delta x}^2_{ \scriptscriptstyle ++}} \right] \!-\! \Bigl(\!
{\scriptscriptstyle ++} \rightarrow {\scriptscriptstyle +-}\!\Bigr) + \frac{i 4 
\pi^{2 - \frac12 \epsilon} \mu^{-\epsilon}}{\Gamma(1 \!-\! \frac{\epsilon}2 )} 
\delta^D(x \!-\! x') \! \right\} . \qquad
\end{eqnarray}
The appropriate local term for $B_3$ is,
\begin{eqnarray}
B_{3L}(x) & = & \frac{-f H^{2 - \epsilon} \mu^{-\epsilon}}{2^{6 - 
\epsilon} \pi^{4-\epsilon}} \frac{ \Gamma(3 \!-\! \epsilon) a^{-3+\epsilon}}{
(1 \!-\! \frac{\epsilon}2) (1 \!-\! \epsilon) \epsilon} \, \Bigl\{\ln(a) 
\partial^2 a + \partial^2 \Bigl(a \ln(a)\Bigr) \Bigr\} \; , \\
& = & \frac{f H^{4 - \epsilon} \mu^{-\epsilon}}{8 \pi^{4- \epsilon} } 
\frac{\Gamma(1 \!-\! \epsilon)}{\epsilon} \Bigl\{\ln(a) + \frac34\Bigr\} 
a^{\epsilon} .
\end{eqnarray}

For the nonlocal terms it is useful to extract a factor to go with 
(\ref{factor}) making a total multiplicative factor of,
\begin{eqnarray}
\lefteqn{- i f H^{2 - \epsilon} \frac{\Gamma(2 \!-\! \frac{\epsilon}2) 
\Gamma(3 \!-\! \epsilon)}{2^{4 - \epsilon} \pi^{6 - \frac32 \epsilon}
a^{3-\epsilon}} 
\times \frac{\mu^{-2\epsilon} a'}{16 (1 \!-\! \frac{\epsilon}2)^2 (1 \!-\!
\epsilon)} } \nonumber \\
& & \hspace{4cm} = - i f H^{2 - \epsilon} \mu^{-2 \epsilon} \frac{\Gamma(1
\!-\! \frac{\epsilon}2) \Gamma(1 \!-\! \epsilon)}{ 2^{7 - \epsilon} \pi^{6 - 
\frac32 \epsilon} a^{3-\epsilon} } \, a' . \qquad \label{newfac}
\end{eqnarray}
It is also useful to note the expansions for the Gamma function,
\begin{equation}
\Gamma(1 \!-\! \epsilon) = 1 + \gamma \epsilon + \Bigl[\frac{\gamma^2}2 +
\frac{\pi^2}{12}\Bigr] \epsilon^2 + \dots \quad \Longrightarrow \quad 
\frac{\Gamma^2(1 \!-\! \frac{\epsilon}2)}{\Gamma(1 \!-\! \epsilon)} = 
1 -\frac{\pi^2}{24} \epsilon^2 + \dots \; ,
\end{equation}
and the cotangent,
\begin{equation}
\frac{\pi}2 \cot\Bigl(\frac{\pi \epsilon}2\Bigr) = \frac1{\epsilon} \left\{
1 - \frac{\pi^2}{12} \epsilon^2 + O(\epsilon^4)\right\} .
\end{equation}
We can expand the nonlocal integrands without regard to $\pm$ variations.

The integrand for the nonlocal part of $B_1(x)$ is (\ref{newfac}) times,
\begin{eqnarray}
\lefteqn{
\frac{(1 \!-\! \frac{\epsilon}4) (1 \!-\! \frac{\epsilon}2)^3}{(1 \!-\! 
\frac34 \epsilon) (1 \!-\! \epsilon) (1 \!-\! \frac32 \epsilon)}
\frac{\Gamma^2(1 \!-\! \frac{\epsilon}2)}{\Gamma(1 \!-\! \epsilon)}
\Bigl(\frac{H^2 a a'}{4 \mu^2}\Bigr)^{\frac{\epsilon}2} \frac{\partial^4}{
\epsilon} \left\{ \frac{\ln(\mu^2 {\Delta x}^2) \!+\! \epsilon \ln^2(\mu^2 
{\Delta x}^2)}{\Delta x^2} \right\} } \nonumber \\
& & \hspace{2cm} = \frac{\partial^4}{\epsilon} \left[ \frac{\ln(\mu^2 
{\Delta x}^2)}{\Delta x^2} \right] \!+\! \frac32 \partial^4 \left[ 
\frac{\ln(\mu^2 {\Delta x}^2)}{\Delta x^2} \right] \!+\! \partial^4 
\left[ \frac{\ln^2(\mu^2 {\Delta x}^2)}{\Delta x^2} \right] \nonumber \\
& & \hspace{5cm} + \frac12 \ln\Bigl(\frac{H^2 a a'}{4 \mu^2}\Bigr) \partial^4 
\left[ \frac{\ln(\mu^2 {\Delta x}^2)}{\Delta x^2} \right] \!+\! O(\epsilon) . 
\qquad \label{B1N}
\end{eqnarray}
The integrand for the nonlocal part of $B_2(x)$ is (\ref{newfac}) times,
\begin{eqnarray}
\lefteqn{- \frac{\pi}2 \cot\Bigl(\frac{\pi \epsilon}2\Bigr) \partial^4
\left\{ \frac{\ln(\mu^2 {\Delta x}^2) \!+\! \frac34 \epsilon \ln^2(\mu^2 
{\Delta x}^2)}{\Delta x^2} \right\} } \nonumber \\
& & \hspace{3cm} = -\frac{\partial^4}{\epsilon} \left[ \frac{\ln(\mu^2 
{\Delta x}^2)}{\Delta x^2} \right] \!-\! \frac34 \partial^4 \left[ 
\frac{\ln^2(\mu^2 {\Delta x}^2)}{\Delta x^2} \right] \!+\! O(\epsilon) . 
\qquad \label{B2N}
\end{eqnarray}
And the integrand for the nonlocal part of $B_3(x)$ is (\ref{newfac}) times,
\begin{equation}
-\frac12 \ln(a a') \partial^4 \left[ \frac{\ln(\mu^2 {\Delta x}^2)}{\Delta x^2}
\right] \!+\! O(\epsilon) . \qquad \label{B3N}
\end{equation}
Combining (\ref{B1N}-\ref{B3N}), multiplying by (\ref{newfac}), including the
integral, and taking $\epsilon = 0$, gives the following result for the 
nonlocal part of the $B$ term,
\begin{eqnarray}
\lefteqn{B_N(x) = -\frac{i f H^2}{2^7 \pi^6} a^{-3} \partial^4 \!\! 
\int d^4x'
a' \left\{\frac{\Bigl[ \frac32 \!+\! \ln(\frac{H}{2 \mu})\Bigr] \ln(\mu^2 
{\Delta x}^2_{ \scriptscriptstyle ++}) \!+\! \frac14 \ln^2(\mu^2 {\Delta x}^2_{
\scriptscriptstyle ++})}{{\Delta x}^2_{\scriptscriptstyle ++}} \right. }
\nonumber \\
& & \hspace{3.5cm}\left. - \frac{\Bigl[ \frac32 \!+\! \ln(\frac{H}{2 \mu})\Bigr]
\ln(\mu^2 {\Delta x}^2_{\scriptscriptstyle +-}) \!+\! \frac14 \ln^2(\mu^2 
{\Delta x}^2_{\scriptscriptstyle +-})}{{\Delta x}^2_{\scriptscriptstyle +-}} 
\right\} . \qquad
\end{eqnarray}

The reduction of $B_N(x)$ proceeds very much like that of $A_N(x)$. We begin
by extracting another d'Alembertian, then combine $++$ and $+-$ parts to
make causality and reality manifest, and perform the angular integrations,
\begin{eqnarray}
\lefteqn{B_N(x) = -\frac{i f H^2}{2^7 \pi^6} a^{-3} \partial^6 \!\! 
\int d^4x' a' } \nonumber \\
& & \hspace{-.4cm} \times \!\! \left\{ \!\! \matrix{\frac1{48} \ln^3(\mu^2 
{\Delta x}^2_{\scriptscriptstyle ++}) \!+\! \frac18 \ln\Bigl(\frac{e H}{2 
\mu}\Bigr) \ln^2(\mu^2 {\Delta x}^2_{\scriptscriptstyle ++}) \!-\! \frac14 
\ln\Bigl(\frac{e H}{2 \mu}\Bigr) \ln(\mu^2 {\Delta x}^2_{\scriptscriptstyle 
++}) \cr
- \frac1{48} \ln^3(\mu^2 {\Delta x}^2_{\scriptscriptstyle +-}) \!-\! \frac18 
\ln\Bigl(\frac{e H}{2 \mu}\Bigr) \ln^2(\mu^2 {\Delta x}^2_{\scriptscriptstyle 
+-}) \!+\! \frac14 \ln\Bigl(\frac{e H}{2 \mu}\Bigr) \ln(\mu^2 {\Delta x}^2_{
\scriptscriptstyle +-})} \!\! \right\} , \qquad \\
& & \hspace{-.4cm} = -\frac{f H^2}{2^6 \pi^4} a^{-3} \partial_0^6 \!\! 
\int_{\eta_I}^{\eta} \!\! d\eta' a' \!\! \int_0^{\Delta \eta} \!\!\!\!\!\!\! 
dr \, r^2 \nonumber \\
& & \hspace{-.4cm} \times \!\! \left\{ \frac14 \ln^2\Bigl[\mu^2 \Bigl({\Delta 
\eta}^2 \!-\! r^2\Bigr)\Bigr] \!-\! \frac{\pi^2}{12} \!+\! \ln\Bigl(\frac{e 
H}{2 \mu}\Bigr) \ln\Bigl[\mu^2 \Bigl({\Delta \eta}^2 \!-\! r^2 \Bigr)\Bigr] 
\!-\! \ln\Bigl(\frac{e H}{2 \mu}\Bigr) \right\} . \qquad
\end{eqnarray}
The next step is to make the change of variables $r = {\Delta \eta} z$ and 
perform the integration over $z$. For this purpose it is useful to note
the integrals,
\begin{equation}
\int_0^1 dz \, z^2 \ln\Bigl(\frac{1 \!-\! z^2}4\Bigr) = -\frac89 \quad 
{\rm and} \quad \int_0^1 dz \, z^2 \ln^2\Bigl(\frac{1 \!-\! z^2}4\Bigr) = 
\frac{104}{27} - \frac{\pi^2}9 \; .
\end{equation}
Applying these identities gives,
\begin{eqnarray}
\lefteqn{B_N(x) = -\frac{f H^2}{2^6 \pi^4} a^{-3} \partial_0^6 \!\! 
\int_{\eta_I}^{\eta} \!\! d\eta' a' {\Delta \eta}^3 \!\! \left\{ \frac13
\ln^2\Bigl(2 \mu {\Delta \eta}\Bigr) \right. } \nonumber \\
& & \hspace{1.5cm} \left. - \frac89 \ln\Bigl(2 \mu {\Delta \eta}\Bigr) \!+\! 
\frac{26}{27} \!-\! \frac{\pi^2}{18} \!+\! \ln\Bigl(\frac{e H}{2 \mu}\Bigr) 
\Bigl[\frac23 \ln\Bigl(2 \mu {\Delta \eta}\Bigr) - \frac{11}9\Bigr] \right\} . 
\qquad
\end{eqnarray}

The next step is to bring three of the derivatives inside,
\begin{eqnarray}
\lefteqn{B_N(x) = -\frac{f H^2}{2^6 \pi^4} a^{-3} \partial_0^5 \!\! 
\int_{\eta_I}^{\eta} \!\! d\eta' a' {\Delta \eta}^2 \left\{
\ln^2\Bigl(2 \mu {\Delta \eta}\Bigr) \right. } \nonumber \\
& & \hspace{2cm} \left. - 2 \ln\Bigl(2 \mu {\Delta \eta}\Bigr) \!+\! 
2 \!-\! \frac{\pi^2}6 \!+\! \ln\Bigl(\frac{e H}{2 \mu}\Bigr) \Bigl[2 
\ln\Bigl(2 \mu {\Delta \eta}\Bigr) - 3\Bigr] \right\} , \qquad \\
& & = -\frac{f H^2}{2^6 \pi^4} a^{-3} \partial_0^4 \!\! 
\int_{\eta_I}^{\eta} \!\! d\eta' a' {\Delta \eta} \left\{
2 \ln^2\Bigl(2 \mu {\Delta \eta}\Bigr) \right. \nonumber \\
& & \hspace{2cm} \left. - 2 \ln\Bigl(2 \mu {\Delta \eta}\Bigr) \!+\! 
2 \!-\! \frac{\pi^2}3 \!+\! \ln\Bigl(\frac{e H}{2 \mu}\Bigr) \Bigl[4
\ln\Bigl(2 \mu {\Delta \eta}\Bigr) - 4\Bigr] \right\} , \qquad \\
& & = -\frac{f H^2}{2^6 \pi^4 a^3} \partial_0^3 \!\! \int_{\eta_I}^{
\eta} \!\!\! d\eta' a' \left\{ 2 \ln^2\Bigl(2 \mu {\Delta \eta}\Bigr) \!+\! 4 
\ln\Bigl(\frac{e^{\frac32} H}{2 \mu}\Bigr) \ln\Bigl(2 \mu {\Delta \eta}\Bigr) 
\!-\! \frac{\pi^2}3 \right\} . \qquad
\end{eqnarray}
Before performing the integral it is best to rearrange the integrand,
\begin{equation}
\ln^2\Bigl(2 \mu {\Delta \eta}\Bigr) \!+\! 2 \ln\Bigl(\frac{e^{\frac32} H}{2 
\mu}\Bigr) \ln\Bigl(2 \mu {\Delta \eta}\Bigr) \!-\! \frac{\pi^2}6 =
\ln^2(H {\Delta \eta}) + 3 \ln(H {\Delta \eta}) + K \; ,
\end{equation}
where the constant $K$ is,
\begin{equation}
K \equiv - \ln^2\Bigl(\frac{2 \mu}{H}\Bigr) + 3 \ln\Bigl(\frac{2 \mu}{H}\Bigr)
- \frac{\pi^2}6 \; .
\end{equation}
The integral gives a complicated result,
\begin{eqnarray}
\lefteqn{B_N(x) = \frac{f H}{2^5 \pi^4} a^{-3} \partial_0^3 \!\! \left\{
-\frac13 \ln^3(a) \!-\! \frac{\pi^2}3 \ln(a) \!+\! 2 \zeta(3) \!-\!
2 \sum_{n=1}^{\infty} \frac{a^{-n}}{n^3} \right. } \nonumber \\
& & \hspace{2cm} - \ln(a) \ln^2\Bigl(1 \!-\! \frac1{a}\Bigr) \!+\! 
2 \ln\Bigl(1 \!-\! \frac1{a}\Bigr) \sum_{n=1}^{\infty} \frac{(1 \!-\! 
\frac1{a})^n}{n^2} \nonumber \\
& & \hspace{2cm} \left. - 2 \sum_{n=1}^{\infty} \frac{(1\!-\!\frac1{a})^n}{n^2}
\!+\! \frac32 \ln^2(a) \!+\! \frac{\pi^2}2 \!-\! 3 \sum_{n=1}^{\infty} 
\frac{a^{-n}}{n^2} \!-\! K \ln(a) \right\} . \qquad
\end{eqnarray}
In acting the derivatives there is no point to keeping any but the
logarithmically enhanced terms,
\begin{eqnarray}
B_N(x) & = & \frac{f H}{2^5 \pi^4} a^{-3} \partial_0^3 \left\{ -\frac13 
\ln^3(a) \!+\! \frac32 \ln^2(a) + O(\ln(a)) \right\} , \\
& = & - \frac{f H^4}{2^4 \pi^4} \left\{\ln^2(a) + O(1) 
\right\} .
\end{eqnarray}

By comparison the $C$ terms are straightforward. The two nonlocal terms
cancel,
\begin{eqnarray}
C_{1N}(x) & = & -\frac{i 3 f H^4}{2^8 \pi^6 a^2} \partial^2 \int d^4x'
a^{\prime 2} \left\{ \frac{\ln(\mu^2 {\Delta x}^2_{\scriptscriptstyle ++})}{
{\Delta x}^2_{\scriptscriptstyle ++}} - \frac{\ln(\mu^2 {\Delta x}^2_{
\scriptscriptstyle +-})}{{\Delta x}^2_{\scriptscriptstyle +-}} \right\} , \\
C_{2N}(x) & = & +\frac{i 3 f H^4}{2^8 \pi^6 a^2} \partial^2 \int d^4x'
a^{\prime 2} \left\{ \frac{\ln(\mu^2 {\Delta x}^2_{\scriptscriptstyle ++})}{
{\Delta x}^2_{\scriptscriptstyle ++}} - \frac{\ln(\mu^2 {\Delta x}^2_{
\scriptscriptstyle +-})}{{\Delta x}^2_{\scriptscriptstyle +-}} \right\} .
\end{eqnarray}
The corresponding local terms are,
\begin{eqnarray}
C_{1L}(x) & = & + \frac{f H^4 \mu^{-2\epsilon}}{2^7 \pi^{4-\epsilon}}
\, \frac{(1\!-\! \frac{\epsilon}2) \Gamma(2 \!-\! \frac{\epsilon}2) \Gamma(4 
\!-\! \frac{\epsilon}2)}{(1 \!-\! \frac32 \epsilon) (1 \!+\! \epsilon)} \,
\frac{a^{2\epsilon}}{\epsilon} \; , \\
C_{2L}(x) & = & - \frac{f H^{4-\epsilon} \mu^{-\epsilon}}{2^{6-\epsilon}
\pi^{4-\epsilon}} \, \frac{(1\!-\! \frac{\epsilon}2) \Gamma(4 \!-\! \epsilon)}{
(1 \!-\! \frac{\epsilon}4) (1 \!-\! \epsilon)} \, \frac{a^{\epsilon}}{
\epsilon} \; .
\end{eqnarray}

\vskip .5cm

\centerline{\bf Acknowledgements}

It is a pleasure to acknowledge joint early work on this project
with P. M. Ho. We are enormously grateful for extensive conversations 
with T. Prokopec and N. C. Tsamis on the problem of generalizing 
Starobinski\u{\i}'s technique beyond scalars with nonderivative 
interactions. And we thank B. Allen, S. P. Martin, L. Parker and T. N. 
Tomaras for discussions and correspondence concerning the effective 
potential. This work was partially supported by NSF grant PHY-0244714 
and by the Institute for Fundamental Theory at the University of Florida.


\begin{thebibliography}{99}

\bibitem{RPW1} R. P. Woodard, ``Quantum Effects during Inflation,'' in
{\it Norman 2003, Quantum field theory under the influence of external
conditions} (Rinton Press, Princeton, 2004) ed. K. A. Milton, pp. 325-330,
astro-ph/0310757.

\bibitem{OW1} V. K. Onemli and R. P. Woodard, Class. Quant. Grav. {\bf 19}
(2002) 4607, gr-qc/0204065.

\bibitem{OW2} V. K. Onemli and R. P. Woodard, Phys. Rev. {\bf D70} (2004) 
107301, gr-qc/0406098.

\bibitem{VF} A. Vilenkin and L. H. Ford, Phys. Rev. {\bf D26} (1982) 1231.

\bibitem{L} A. D. Linde, Phys. Lett. {\bf B116} (1982) 335.

\bibitem{S} A. A. Starobinski\u{\i}, Phys. Lett. {\bf B117} (1982) 175.

\bibitem{BOW} T. Brunier, V. K. Onemli and R. P. Woodard, Class. Quant. Grav. 
{\bf 22} (2005) 59, gr-qc/0408080.

\bibitem{PTW1} T. Prokopec, O. Tornkvist and R. P. Woodard, Phys. Rev. Lett.
{\bf 89} (2002) 101301, astro-ph/0205331.

\bibitem{PTW2} T. Prokopec, O. Tornkvist and R. P. Woodard, Ann. Phys.
{\bf 303} (2003) 251, gr-qc/0205130.

\bibitem{PW1} T. Prokopec and R. P. Woodard, Am. J. Phys. {\bf 72} (2004) 60, 
astro-ph/0303358.

\bibitem{PW2} T. Prokopec and R. P. Woodard, Ann. Phys. {\bf 312} (2004) 1, 
gr-qc/0310056.

\bibitem{PW3} T. Prokopec and R. P. Woodard, JHEP {\bf 0310} (2003) 059, 
astro-ph/0309593.

\bibitem{GP} B. Garbrecht and T. Prokopec, Phys. Rev. {\bf D73} (2006)
064036, gr-qc/0602011.

\bibitem{TW1} N. C. Tsamis and R. P. Woodard, Ann. Phys. {\bf 238} (1995) 1.

\bibitem{TW2} N. C. Tsamis and R. P. Woodard, Nucl. Phys. {\bf B474} (1996) 
235, hep-ph/9602315.
 
\bibitem{TW3} N. C. Tsamis and R. P. Woodard, Ann. Phys. {\bf 253} (1997) 1,
hep-ph/9602316.

\bibitem{SW} S. Weinberg, Phys. Rev. {\bf D72} (2005) 043514, hep-th/0506236.

\bibitem{MW1} S. P. Miao and R. P. Woodard, Class. Quant. Grav. {\bf 23}
(2006) 1721, gr-qc/0511140.

\bibitem{MW2} S. P. Miao and R. P. Woodard, Phys. Rev. {\bf D74} (2006)
024021, gr-qc/0603135.

\bibitem{AAS} A. A. Starobinski\u{\i}, ``Stochastic de Sitter (inflationary)
stage in the early universe,'' in {\it Field Theory, Quantum Gravity and
Strings}, ed. H. J. de Vega and N. Sanchez (Springer-Verlag, Berlin, 1986)
pp. 107-126.

\bibitem{SY} A. A. Starobinski\u{\i} and J. Yokoyama, Phys. Rev. {\bf D50}
(1994) 6357, astro-ph/9407016.

\bibitem{RPW2} R. P. Woodard, Nucl. Phys. Proc. Suppl. {\bf 148} (2005), 108
astro-ph/0502556.

\bibitem{TW4} N. C. Tsamis and R. P. Woodard, Nucl. Phys. {\bf B724} (2005)
295, gr-qc/0505115.

\bibitem{YF} C. N. Yang and D. Feldman, Phys. Rev. {\bf 79} (1950) 972.

\bibitem{KW} E. O. Kahya and R. P. Woodard, Phys. Rev. {\bf D72} (2205) 104001,
gr-qc/0508015.

\bibitem{TW5} N. C. Tsamis and R. P. Woodard, Annals Phys. {\bf 321}
(2006) 875, gr-qc/0506056.

\bibitem{HAW} H. A. Weldon, Phys. Rev. {\bf D63} (2001) 104010, gr-qc/0009086.

\bibitem{RPW3} R. P. Woodard, Phys. Lett. {\bf B148} (1984) 440.

\bibitem{AF} B. Allen and A. Folacci, Phys. Rev. {\bf D35} (1987) 3771.

\bibitem{BA} B. Allen, Phys. Rev. {\bf D32} (1985) 3136.

\bibitem{CR} P. Candelas and D. J. Raine, Phys. Rev. {\bf D12} (1975) 965.

\bibitem{GR} I. S. Gradshteyn and I. M. Ryzhik, {\it Table of Integrals,
Series and Products}, 4th ed. (Academic Press, New York, 1965).

\bibitem{IMM} T. Inagaki, S. Mukaigawa and T. Muta, Phys. Rev. {\bf D52} 
(1995) 4267, hep-th/9505058.

\bibitem{IMO} T. Inagaki, T. Muta and S. D. Odintsov, Prog. Theor. Phys.
Suppl. {\bf 127} (1997) 93, hep-th/9711084.

\bibitem{DW} L. D. Duffy and R. P. Woodard, Phys. Rev. {\bf D72} (2005) 024023,
hep-ph/0505156.

\bibitem{CW} S. R. Coleman and E. Weinberg, Phys. Rev. {\bf D7} (1973) 1888.

\bibitem{SHR} P. M. Stevenson, G. A. Hajj and J. F. Reed, Phys. Rev. {\bf D34}
(1986) 3117.

\bibitem{FJJ} C. Ford, I. Jack and D. R. T. Jones, Nucl. Phys. {\bf B387} 
(1992) 373, Erratum ibid. {\bf B504} (1997) 551, hep-ph/0111190.

\bibitem{JS} J. Schwinger, J. Math. Phys. {\bf 2} (1961) 407.

\bibitem{M} K. T. Mahanthappa, Phys. Rev. {\bf 126} (1962) 329.

\bibitem{BM} P. M. Bakshi and K. T. Mahanthappa, J. Math. Phys. {\bf 4} (1963)
1; J. Math. Phys. {\bf 4} (1963) 12.

\bibitem{K} L. V. Keldysh, Sov. Phys. JETP {\bf 20} (1965) 1018.

\bibitem{CSHY} K. C. Chou, Z. B. Su, B. L. Hao and L. Yu, Phys. Rept. {\bf 118} 
(1985) 1.

\bibitem{J} R. D. Jordan, Phys. Rev. {\bf D33} (1986) 444.

\bibitem{CH} E. Calzetta and B. L. Hu, Phys. Rev. {\bf D35} (1987) 495.

\bibitem{FW} L. H. Ford and R. P. Woodard, Class. Quant. Grav. {\bf 22} (2005)
1637, gr-qc/0411003.

\bibitem{RPW4} R. P. Woodard, Lect. Notes Phys. {\bf 720} (2007) 403,
astro-ph/0601672.

\bibitem{PP} T. Prokopec and E. Puchwein, JCAP {\bf 0404} (2004) 007,
astro-ph/\-0312274.

\bibitem{PTsW1} T. Prokopec, N. C. Tsamis and R. P. Woodard, Annals 
Phys. {\bf 323} (2008) 1324, arXiv:0707.0847.

\bibitem{PTsW2} T. Prokopec, N. C. Tsamis and R. P. Woodard, Class.
Quant. Grav. {\bf 24} (2007) 201, gr-qc/0607094.

\bibitem{PTsW3} T. Prokopec, N. C. Tsamis and R. P. Woodard, ``Two loop
stress-energy tensor for inflationary scalar electrodynamics,'' 
arXiv:0802.3673.

\end{thebibliography}
\end{document}